\documentclass[12pt]{article}
\usepackage{epsfig,amssymb,amsmath}
\usepackage{jheppub}
\usepackage{soul} 
\usepackage{hyperref}
\usepackage{esint} 
\usepackage{breqn}
\def \e  {\mathop{\rm e}\nolimits}

\newcommand \widebar [1] {\overline{#1}}

\newcommand \vev [1] {\langle{#1}\rangle}

\newcommand \ket [1] {|{#1}\rangle}
\newcommand \bra [1] {\langle {#1}|}
\newcommand\re[1]{(\ref{#1})}
\def \qqquad {\qquad\quad}
\def \qqqquad {\qquad\qquad}

\newcommand{\p}[1]{(\ref{#1})}

\newcommand{\q}{\theta}
\newcommand{\bq}{\bar\theta}

\def\numberbysection{\@addtoreset{equation}{section}
                     \def\theequation{\thesection.\arabic{equation}}}

\date{\today}

\preprint{\small  \parbox[t]{42mm}{IPhT-T15/085 \\ LAPTH-022/15\\CERN-PH-TH-2015-102}}

\title{\Large Four-point correlation function of stress-energy tensors in \boldmath $\mathcal N=4$ superconformal theories}

\author[a]{G. P. Korchemsky}
\author[b,c,d]{and E. Sokatchev}

\affiliation[a]{Institut de Physique Th\'eorique\footnote{Unit\'e Mixte de Recherche 3681 du CNRS}, Universit\'e Paris Saclay, CNRS, CEA, F-91191 Gif-sur-Yvette}

\affiliation[b]{Physics Department, Theory Unit, CERN,
CH -1211, Geneva 23, Switzerland}

\affiliation[c]{Institut Universitaire de France, 103, bd Saint-Michel F-75005 Paris, France}

\affiliation[d]{LAPTH \footnote{Unit\'e Mixte de Recherche 5108 du CNRS, associ\'ee \`a l'Universit\'e de Savoie},   Universit\'{e} de Savoie, CNRS B.P. 110,  F-74941 Annecy-le-Vieux, France}
 
\abstract{We derive the explicit expression for the four-point correlation function of stress-energy
tensors in {four-dimensional} $\mathcal N=4$ superconformal theory. We show that it has a remarkably simple
and suggestive form allowing us to predict a large class of four-point correlation 
functions involving the stress-energy tensor and other conserved currents. We then apply the obtained results on 
the correlation functions to computing the energy-energy correlations, which  measure the flow of energy in the final states 
created from the vacuum by a source. We demonstrate that they are given by a universal function independent
of the choice of the source. Our analysis relies only on $\mathcal N=4$ superconformal symmetry and does 
not use the dynamics of the theory.  } 
  
\begin{document}

\maketitle

\section{Introduction}
 
The  correlation functions of stress-energy tensors are very natural quantities to study in any conformal 
field theory. It is well known that the two-point function is fixed by conformal 
symmetry whereas the higher point correlation functions can have a very complicated form. Namely, they are given by a linear combination of numerous Lorentz structures arising as solutions to the conformal 
Ward identities consistent with the conservation of the stress-energy tensor. Their total number depends both on the 
number of points and the dimension of the space-time. In particular, in a four-dimensional conformal field theory, the three-point 
correlation function involves three different Lorentz structures \cite{Stanev:1988ft,Osborn:1993cr,Costa:2011mg}. They coincide with the three-point correlation functions of stress-energy 
tensors in a free theory of scalars, fermions and gauge fields, respectively.
For four-point functions, the situation is much more complicated -- not only the number of Lorentz structures increases to $22$,
but each of them involves some function of the conformal cross-ratios  \cite{Dymarsky:2013wla}. This makes the calculation of the four-point correlation function of
stress-energy tensors an extremely difficult task. To the best of our knowledge, there exists no closed expression for such correlation function in the literature.

The problem becomes significantly simpler  if the underlying conformal theory has supersymmetry. In general, it leads to additional constraints on the correlation functions of the stress-energy tensors and, therefore, greatly reduces the number of independent
functions of cross-ratios.~\footnote{For some general results on superconformal correlation functions see, e.g., \cite{Howe:1996rk, Park:1999pd, Eden:1999gh, Dolan:2001tt, Heslop:2002hp, Dolan:2004mu}.}
In the four-dimensional maximally supersymmetric  $\mathcal N=4$  
theory this number shrinks to one, so that the four-point correlation of stress-energy tensors depends on a single scalar
function. The same function determines the four-point correlation function of the scalar $1/2$ BPS operators $O_{\boldsymbol{20'}}$.
The reason for this is that the two operators, $O_{\boldsymbol{20'}}$ and $T_{\mu\nu}$, belong to the same  $\mathcal N=4$
supermultiplet, the so-called stress-energy supermultiplet $\mathcal T$, and their correlation functions are related to each other by  
$\mathcal N=4$ superconformal Ward identities. The general solution to these identities, defining the four-point correlation function
of the supercurrents $\vev{\mathcal T(1) \dots \mathcal T(4)}$, was derived in \cite{Belitsky:2014zha}. Different correlation functions involving
$O_{\boldsymbol{20'}}$ and $T_{\mu\nu}$  appear as its components.

In this paper, we derive the explicit expression for the four-point correlation function of stress-energy
tensors in an $\mathcal N=4$ superconformal theory. We show that it has a remarkably simple
and suggestive form allowing us to extend the obtained results to a larger class of four-point correlation functions involving the 
stress-energy tensor and other conserved currents. Our analysis relies only on $\mathcal N=4$ superconformal symmetry and
does not use the dynamics of the theory. In the special case of $\mathcal N=4$ supersymmetric Yang-Mills
theory (SYM), the four-point correlation functions are determined by the scalar function $\Phi(u,v)$, which has been extensively studied at weak
and at strong coupling.
 
Among all four-point correlation functions described by $\vev{\mathcal T(1) \mathcal T(2) \mathcal T(3) \mathcal T(4)}$, those involving 
two stress-energy  tensors play a special role. They can be used to compute interesting (infrared safe) observables, the so-called
energy-energy correlations \cite{Basham:1978bw}, measuring the flow of energy in the final states created from the vacuum by some source $J(x)$ \cite{Ore:1979ry,Sveshnikov:1995vi,Korchemsky:1997sy,Korchemsky:1999kt,Belitsky:2001ij}. The choice of the source is arbitrary and physically interesting cases involve the 
1/2 BPS operator $O_{\bf 20'}$, the stress-energy tensor, the
$R-$symmetry current and the Lagrangian of the theory, all belonging to the stress-energy supermultiplet $\mathcal T$.
{The last two cases are of particular interest since they can be thought of as prototypes of the electromagnetic current in QCD and of the effective coupling
of Higgs boson to gluons in the Standard Model, respectively.}

The energy-energy correlations can be obtained from the four-point correlation
function $\vev{0|T_{\mu_1\nu_1}(1)T_{\mu_2\nu_2}(2)J(3)J(4)|0}$ through a limiting procedure described in great detail in \cite{Belitsky:2013xxa,Belitsky:2013bja}.
Using this approach, the energy-energy correlations have been computed in $\mathcal N=4$ SYM theory 
for a scalar source $J=O_{\bf 20'}$ 
at next-to-leading order both at weak and at strong coupling \cite{Hofman:2008ar,Belitsky:2013ofa,Goncalves:2014ffa}.
Based on general physical considerations, we would expect that the flow of energy in the final state should depend on the choice
of the source. Using the obtained
results on the correlation functions, we find that, quite surprisingly, 
the energy-energy correlations  in $\mathcal N=4$ superconformal   theory are the same for the different  source operators mentioned above and, 
therefore, are universal.

The paper is organized as follows. In section \ref{sect:2} we describe the properties of the stress-energy supermultiplet
and present the general expression for the four-point correlation function   satisfying the $\mathcal N=4$ superconformal Ward identities.
In section \ref{sect:TTTT} we explain how to extract the four-point correlation function of stress-energy tensors from the supercorrelator
 $\vev{\mathcal T(1) \mathcal T(2) \mathcal T(3) \mathcal T(4)}$. Then we use the special properties of the $\mathcal N=4$ superconformal   generators 
 to derive its explicit expression. In section \ref{sect:curr} we extend the analysis to correlation functions involving conserved currents.
 We argue that they have  a remarkably simple universal form and illustrate this by a few examples.
 In section \ref{sect:app} we apply the obtained expressions for the correlation functions to the evaluation of the energy-energy correlations for
 different source operators. Section \ref{sect:conc} contains concluding remarks. Two appendices contain technical details.
 
\section{Superconformal Ward identities}\label{sect:2}

In this section, we summarize the properties of the correlation functions of the $\mathcal N=4$ stress-energy supermultiplet.  
As was already mentioned, this supermultiplet includes the 1/2 BPS scalar operator $O_{\bf 20'}$,  all the conserved currents and the Lagrangian 
of the theory. These operators appear as various components in the expansion of the supercurrent $\mathcal T$ 
in powers of  the Grassmann variables.

\subsection{Stress-energy supermultiplet}

For $\theta=\bar\theta=0$, the lowest component of $\mathcal T$
is the scalar operator $O_{\bf 20'}$. It has conformal weight $2$, belongs to the 
representation $\bf 20'$ of the $R-$symmetry group $SU(4)$ and has the following form
\begin{align}\label{O20}
O_{\bf 20'}(x,y) = O^{AB,CD}(x) Y_{AB} Y_{CD}=\mathcal T(x,\theta,\bar\theta,y)\Big|_{\theta=\bar\theta=0}\,,
\end{align}
where the $SU(4)$ indices take values, e.g., $A=1,\dots,4$.
The auxiliary tensors $Y_{AB} Y_{CD}$ have been introduced to project the operator $O^{AB,CD}(x)$ onto the representation
$\bf 20'$. They satisfy the relations $Y_{AB}=-Y_{BA}$ and $\epsilon^{ABCD} Y_{AB} Y_{CD}=0$ and can be parameterized
as 
\begin{align}\label{Y-proj}
Y_{AB} = u_A^{+c} \epsilon_{cd} u_B^{+d} = \left[\begin{array}{cl} \epsilon_{ab} & -y_{ab'} \\ y_{ba'} & y^2\epsilon_{a'b'}  \end{array}\right],
\end{align}
where $y^2=\det \|y_{aa'} \| = \frac12 y_{aa'} y_{bb'}\epsilon^{ab}\epsilon^{a'b'}$ and we used composite indices $A=(a,a')$ with $a,a'=1,2$ and similarly for $B=(b,b')$. The variables $y_{aa'}$ 
 have the meaning of coordinates of the operator $O_{\bf 20'}(x,y)$ on the four-dimensional coset  $SU(4)/(SU(2)\times SU(2)'\times U(1))$ of the $R-$symmetry group.  
 
In addition, the supercurrent $\mathcal T$ describes a short supermultiplet of the 1/2 BPS type. This means that it is annihilated by half of the super-Poincar\'e generators and, therefore, it effectively depends on $4$ 
chiral and $4$ antichiral Grassmann variables, $\theta_\alpha^a$ and $\bar\theta^{\dot\alpha}_{a'}$  (with $\alpha,\dot\alpha=1,2$ and 
$a,a'=1,2$), respectively. The stress-energy tensor can be obtained by applying $\mathcal N=4$ supersymmetry  
transformations to \re{O20} and can be extracted from $\mathcal T$ with the help of the differential operator~\cite{Belitsky:2014zha}
\begin{align}\notag\label{T-diff.op}
T_{\alpha\dot\alpha,\beta\dot\beta}(x) = \Big[-(\partial_\theta)_\alpha^a (\partial_\theta)_{\beta a} (\partial_{\bar\theta})_{\dot\alpha a'}(\partial_{\bar\theta})_{\dot\beta}^{a'} -(\partial_\theta)^a_{(\alpha} (\partial_x)_{\beta)(\dot\beta} (\partial_y)_{aa'} (\partial_{\bar\theta})_{\dot\alpha)}^{a'}{}&
 \\
 +\frac16 (\partial_x)_{(\alpha\dot\alpha}(\partial_x)_{\beta)\dot\beta} (\partial_y)_{aa'} (\partial_y)^{a'a}\Big]
\mathcal T(x,\theta,\bar\theta,y){}& \Big|_{\theta=\bar\theta=0}\,,
\end{align}
where we switched to spinor notation, $T_{\alpha\dot\alpha,\beta\dot\beta}=\sigma^\mu_{\alpha\dot\alpha}\sigma^\nu_{\beta\dot\beta}T_{\mu\nu}$ (see Appendix \ref{App-conv} for details) and denoted the weighted symmetrization of the indices by, e.g., $A_{(\alpha\beta)}= \frac12 (A_{\alpha\beta}+ A_{\beta\alpha})$. The first term in the  brackets in \re{T-diff.op} selects the $(\theta^2)_{\alpha\beta}(\bar\theta^2)_{\dot\alpha\dot\beta}$ component of the supercurrent, whereas the second and third terms involve total derivatives acting on lower components of $\mathcal T$. These are typical conformal and $R-$symmetry descendant terms, which have to be subtracted in order  to ensure the expected properties of the stress-energy tensor: current conservation $(\partial_x)^{\dot\alpha\alpha} T_{\alpha\dot\alpha,\beta\dot\beta}(x) =0$ and zero $R-$charge or, equivalently,
independence on the $y-$variables, $(\partial_y)^{a'a} T_{\alpha\dot\alpha,\beta\dot\beta}(x) =0$.

In a similar manner, the $R-$symmetry current is related to the supercurrent $\mathcal T$ by the following differential operator \cite{Belitsky:2014zha}
\begin{align}\label{J}
J_{\alpha\dot\alpha,aa'}(x,y) = \Big[ (\partial_{\bar\theta})_{\dot\alpha a'}(\partial_\theta)_{\alpha a} + \frac12 (\partial_x)_{\alpha\dot\alpha}
(\partial_y)_{aa'}\Big]
\mathcal T(x,\theta,\bar\theta,y)\Big|_{\theta=\bar\theta=0}\,,
\end{align}
where the second term in the  brackets is again the subtraction of a descendant needed to ensure the current conservation, $(\partial_x)^{\dot\alpha\alpha}J_{\alpha\dot\alpha,aa'}(x,y) =0$.

\subsection{Auxiliary spinor variables}  
   
Let us start with the two- and three-point correlation functions of the supercurrent $\mathcal T(i)\equiv \mathcal T(x_i,\theta_i,\bar\theta_i,y_i)$. The  $\mathcal N=4$ superconformal symmetry fixes them up to a normalization constant \cite{23pts1,23pts2,23pts3,23pts4,23pts5}
\begin{align}\notag\label{2pt}
  \vev{\mathcal T(1) \mathcal T(2)} {}& =  \frac{c}2 (D_{12})^2\,,
\\
  \vev{\mathcal T(1) \mathcal T(2)\mathcal T(3)} {}& = c\, D_{12} D_{23} D_{31}\,,
\end{align}   
where $D_{ij}$ is a free (super) propagator
\begin{align}\label{D}
D_{ij} = {y_{ij}^2\over \hat x_{ij}^2} =  {\hat y_{ij}^2\over  x_{ij}^2}\,, 
\end{align}
with $\hat x_{ij}^{\dot\alpha\alpha} = x_{ij}^{\dot\alpha\alpha} + \bar\theta_{ij}^{\dot\alpha a'}(y_{ij}^{-1})_{a'a} \theta_{ij}^{\alpha a}    $ and $\hat y_{ij}^{aa'} = y_{ij}^{aa'} +  \theta_{ij}^{\alpha a}  (x_{ij}^{-1})_{\alpha\dot\alpha} \bar\theta_{ij}^{\dot\alpha a'}  $ for
$x_{ij}=x_i-x_j$ and similarly for $y_{ij}$, $\theta_{ij}$ and $\bar\theta_{ij}$ (see Appendix \ref{App-conv} for our conventions for
rising and lowering Lorentz and the $SU(2)$ indices).
In the $\mathcal N=4$ SYM theory with the $SU(N)$ gauge group we have $c=N^2-1$.

Putting $\theta=\bar\theta=0$ on both sides of \re{2pt} we reproduce the known expressions for the two- and three-point correlation functions of the 1/2 BPS operators \re{O20}.
To obtain the correlation functions of the stress-energy tensor  from \re{2pt}, we have to apply the differential operator 
defined in \re{T-diff.op} to each $\mathcal T(i)$ in \re{2pt}. After a  lengthy calculation we arrive at the expected
result for the two-point correlation function (with $x_{12}=x_1-x_2$)
\begin{align}\label{TT}
\vev{T_{\alpha_1 \dot\alpha_1,\beta_1\dot\beta_1}(x_1) T_{\alpha_2 \dot\alpha_2,\beta_2\dot\beta_2}(x_2)} =
160 \,c { (x_{12})_{(\alpha_1 \dot\alpha_2}(x_{12})_{\beta_1) \dot\beta_2}(x_{12})_{(\alpha_2 \dot\alpha_1}(x_{12})_{\beta_2) \dot\beta_1}  \over (x_{12}^2)^6}\,.
\end{align}
Likewise, we can use the second relation in \re{2pt} to reproduce the known result for 
the three-point correlation function of stress-energy tensors in an $\mathcal N=4$ superconformal theory \cite{Dolan:2001tt}.
 
We can simplify  \re{TT} by contracting the Lorentz indices of the stress-energy tensors  
with auxiliary light-like vectors $n_i^\mu$ (with $n_i^2=0$) 
\begin{align}\label{T-pro}
T(i) \equiv n_i^\mu n_i^\nu \,T_{\mu\nu} (x_i) = \frac14 
\lambda_i^\alpha \bar\lambda_i^{\dot\alpha}\lambda_i^\beta \bar\lambda_i^{\dot\beta}\,T_{\alpha\dot\alpha,\beta\dot\beta}(x_i) \,,
\end{align}
where $n_i^{\dot\alpha\alpha} = n_i^\mu (\tilde\sigma_\mu)^{\dot\alpha\alpha} =\lambda_i^\alpha \bar\lambda_i^{\dot\alpha}$ with
$\lambda_i$ and $\bar\lambda_i$ being arbitrary two-component (anti)chiral commuting spinors.
Then, we find from \re{TT}
\begin{align}\label{TT-new}
\vev{T(1) T(2)} = 160\, c\, {\big(\bra{1} x_{12}|2] \bra{2} x_{21}|1]\big)^2 \over (x_{12}^2)^6}\,,
\end{align}
where we used the standard notation for $\bra{i} x_{ij} |j]\equiv \lambda_i^\alpha (x_{ij})_{\alpha\dot\alpha} 
\bar\lambda_j^{\dot\alpha}$. 

The auxiliary variables $\lambda_i$ and $\bar\lambda_i$ serve two main purposes. 
They automatically symmetrize the correlation function with respect to the chiral
and antichiral Lorentz indices and, in addition, they simplify the conformal properties of the stress-energy tensor.
To see this, we exploit the freedom in defining $\lambda_i$ and $\bar\lambda_i$ to assign to them weights under conformal  
inversion $I[(x_i)_{\alpha\dot\alpha}] = (x_i^{-1})_{\alpha\dot\alpha}$. Then, we choose \cite{Drummond:2008vq}
\begin{align}\label{inv-la}
%I[(x_i)_{\alpha\dot\alpha}] = (x_i^{-1})_{\alpha\dot\alpha}\,,\qquad 
I[\lambda_i^\alpha] = w(x_i) (x_i^{-1})^{\dot\alpha\beta} \lambda_{i\beta}\,,\qqqquad 
I[\bar\lambda_i^{\dot\alpha}] = w(x_i) \bar\lambda_{i\dot\beta}(x_i^{-1})^{\dot\beta\alpha}\,,
\end{align}
with a scalar weight factor $w(x_i)=(x_i^2)^n$ that can be chosen at our convenience.  In this section, we put  $w(x_i)=1$
for simplicity but, as shown in section~\ref{sect:ans}, the choice of $w(x_i)=x_i^2$ is more convenient for discussing the 
conformal properties of the four-point correlation functions. 

We would like to mention that the variables $\lambda_i$ and $\bar\lambda_i$ have a natural interpretation
in the context of scattering amplitudes where they are used to describe the helicity states of scattered massless particles. The invariance
of the scattering amplitudes in planar $\mathcal N=4$ SYM under transformations \re{inv-la}  led to the discovery of the
dual conformal symmetry in this theory  \cite{Drummond:2008vq}. As we will see in the next section, the final expression for 
the four-point correlation function of stress-energy tensors involves the same $\mathcal N=4$ dual conformal invariants 
that have been encountered in the study of the scattering amplitudes.

With the assignment   \re{inv-la} and with  $w(x_i)=1$ 
the quantity $\bra{i} x_{ij} |j]$ entering \re{TT-new} transforms covariantly under inversion, 
$I\big[\bra{i} x_{ij} |j] \big] = \bra{i} x_{ij} |j]/(x_i^2 x_j^2)$ leading to the simple transformation property of the two-point 
correlation function \re{TT-new}
\begin{align}
I \big[ \vev{T(1) T(2)} \big] = (x_1^2 x_2^2)^2 \vev{T(1) T(2)}\,.
\end{align}
We deduce from this relation that the projected stress-energy tensor \re{T-pro} transforms under conformal transformations
as a scalar conformal primary operator with conformal weight $2$, 
$I \big[  T(i)  \big] = (x_i^2)^2 T(i)$ \cite{Giombi:2011rz,Todorov:2012xx}. Notice that, by
definition, $T(i)$ is a homogenous polynomial in $\lambda_i$ and $\bar\lambda_i$ of degree $2$. These two facts will 
play an important role in what follows.
   
\subsection{Four-point correlation functions} 

In distinction with \re{2pt}, the four-point correlation function of the supercurrents is not fixed by 
$\mathcal N=4$ superconformal symmetry. The corresponding Ward identities imply that this correlation function can be decomposed into a sum of two terms carrying different information about the states propagating in various OPE channels
\begin{align}\label{G4}
\mathcal G_4= \vev{\mathcal T(1)\mathcal T(2)\mathcal T(3)\mathcal T(4) } =\mathcal G_4^{(0)} + \mathcal G_4^{(\rm anom)},
\end{align} 
Here $\mathcal G_4^{(0)}$ receives
contributions only  from the protected operators in the theory whereas all unprotected operators contribute to $\mathcal G_4^{(\rm anom)}$. As a consequence,  the form $\mathcal G_4^{(0)}$ is fixed by $\mathcal N=4$ superconformal symmetry and is given by an expression
analogous to \re{2pt}
\begin{align}\notag\label{G4-rat}
\mathcal G_4^{(0)} =  c ( D_{12} D_{23} D_{34} D_{41} + D_{13} D_{23} D_{24} D_{14} + D_{12} D_{24} D_{34} D_{13} )
\\
 + \frac{c^2}4 (D_{12}^2 D_{34}^2 +D_{13}^2 D_{24}^2 +D_{14}^2 D_{23}^2 ),
\end{align}
where the second line corresponds to the disconnected contribution. 

The second term $\mathcal G_4^{(\rm anom)}$ has more complicated form and, in contrast with \re{G4-rat}, it is not a rational function of the distances $x_{ij}^2$. It admits the following representation
\begin{align}\label{G4-anom}
\mathcal G_4^{(\rm anom)} {}& =  (D_{13} D_{24})^2\,\mathcal I_4(x,\theta,\bar\theta,y)\,,
\end{align}
where the propagators  $D_{ij}$ were defined in \re{D} and $\mathcal I_4$ depends on the four superspace points. The product of propagators $(D_{13} D_{24})^2$ carries the (super)conformal weights of the supercurrents whereas $\mathcal I_4$ is invariant under $\mathcal N=4$ superconformal transformations, $\mathcal J \,\mathcal I_4=0$ with the generators
$\mathcal J=\{ Q_\alpha^A, \bar S_{\dot\alpha}^A, \bar Q_{\dot\alpha}^A, S_{\alpha}^A, \dots\}$ given in \re{gens} below.

As was shown in \cite{Belitsky:2014zha}, the invariant $\mathcal I_4$ admits two equivalent representations
\begin{align}\notag\label{I4}
\mathcal I_4  {}& =   Q^8 \bar S^8 \left[ \theta_1^4\, \theta_2^4\, \theta_3^4\, \theta_4^4 \,{F(x)\over (D_{13} D_{24})^{2} }\right]
\\ 
{}&=   \bar Q^8  S^8 \left[ \bar\theta_1^4\,\bar \theta_2^4\, \bar\theta_3^4\, \bar\theta_4^4 \,{F(x)\over (D_{13} D_{24})^{2} }\right],
\end{align}
where $Q^8=\prod_{\alpha,A} Q_\alpha^A$ and $\bar S^8=\prod_{\dot\alpha,A}\bar S_{\dot\alpha}^A$ denote the products of the $8$ chiral Poincar\'e supercharges and the $8$ generators of antichiral special superconformal transformations, respectively, and similarly for the generators of opposite chirality. Since the generators $Q$ and $\bar S$ are nilpotent and form an abelian subalgebra, $\{Q,\bar S\}=0$, it immediately
follows from the first relation in \re{I4} that $\mathcal I_4$ is annihilated by these generators. Similarly, it follows from the
second relation in \re{I4} that $\mathcal I_4$ is also annihilated by $\bar Q$ and $S$. 

The expression in the  brackets in \re{I4} contains the product of all the available (anti) chiral Grassmann variables, 
$\theta_i^4=\prod_{\alpha,a} \theta_i^{\alpha a}$ and $\bar \theta_i^4=\prod_{\dot\alpha,a'} \bar\theta_i^{\dot\alpha a'}$,  as well as a scalar 
function $F$ depending only on  the coordinates $x_i$. The Bose symmetry of \re{G4-anom} implies that $F(x)$ should
be invariant under the exchange of any pair of points $x_i\leftrightarrow x_j$. In addition, the invariance of \re{I4} under conformal 
transformations leads to the following inversion property of  $F$ \cite{Belitsky:2014zha}
\begin{align}\label{inv-F}
I\left[ F(x)\right] = (x_1^2 x_2^2 x_3^2 x_4^2)^4 F(x)\,.
\end{align}
It allows us to express $F(x)$ in terms of a function of conformal cross-ratios  
%$u=x_{12}^2 x_{34}^2/(x_{13}^2 x_{24}^2)$ and $v=x_{23}^2 x_{14}^2/(x_{13}^2 x_{24}^2)$
\begin{align}\label{F-Phi}
F(x) = {\Phi(u,v) \over u v (x_{13}^2 x_{24}^2)^4 }\,,\qqquad u={x_{12}^2 x_{34}^2\over x_{13}^2 x_{24}^2}
,\qqquad v={x_{23}^2 x_{14}^2\over x_{13}^2 x_{24}^2}\,.
\end{align} 
The symmetry of $F(x)$ under exchange of points translates into the crossing symmetry relations 
\begin{align}
\Phi(u,v)=\Phi(v,u)={1\over v} \Phi\left({u\over v},{1\over v}\right).
\end{align}

Substituting \re{I4} and \re{F-Phi} into \re{G4-anom} we find that the general expression for  $\mathcal G_4^{(\rm anom)}$ depends on a single function $\Phi(u,v)$. The same applies to the variety of four-point correlation functions that appear as components
in the expansion of $\mathcal G_4^{(\rm anom)}$ in powers of the Grassmann variables. In particular,
putting $\theta_i=\bar\theta_i=0$ on both sides of \re{G4-anom} we find that the function $\Phi(u,v)$ defines the four-point correlation 
function of 1/2 BPS operators $O_{\bf 20'}$. The latter correlation function has been thoroughly studied in $\mathcal N=4$ SYM 
\cite{Eden:2011we,Eden:2012tu}
and the function $\Phi(u,v)$ is known in this theory both at weak \cite{GonzalezRey:1998tk,Eden:2000mv,Bianchi:2000hn,Drummond:2013nda} and at strong coupling \cite{Arutyunov:1999fb,Arutyunov:2000py,Arutyunov:2000ku}.
  
\subsection{Master formula}   

Let us simplify relation \re{G4-anom}.
To begin with, we rewrite it using the first relation in \re{I4}
\begin{align}\label{G4-start}
\mathcal G_4^{(\rm anom)} {}& =  (D_{13} D_{24})^2\, Q^4 Q'{}^4  \bar S^4\bar S'{}^4 \left[ \theta_1^4\, \theta_2^4\, \theta_3^4\, \theta_4^4 \,{F(x)\over (D_{13} D_{24})^{2}} \right],
\end{align}
where we used the composite index $A=(a,a')$ to split the product of $8$ generators $Q_\alpha^A$  into $Q^4=\prod_{\alpha,a} 
Q_{\alpha}^a$ and $Q'{}^4=\prod_{\alpha,a'} Q_\alpha^{a'}$ and similarly for the product of  generators $\bar S_{\dot\alpha}^a$ and $\bar S_{\dot\alpha}^{a'}$.
These generators   act on the superspace coordinates $(x_i,\q_i,\bq_i,y_i)$ (with $i=1,\dots,4$) at each point and
are given by the sum over the 4 points 
of the following anticommuting differential operators 
\begin{align}\notag\label{gens}
& Q^\alpha_a = {\partial \over \partial \q_\alpha^a}\,,
\\\notag
& Q^\alpha_{a'} = \bq_{a'\dot\alpha} {\partial\over\partial x_{\alpha\dot\alpha}} + y_{a'}{}^a {\partial\over\partial \theta_{\alpha}^a}\,,
\\\notag
& \bar S_{a\dot\beta} = -\bq_{a'\dot\beta} {\partial \over \partial y_{a'}{}^a}  +   {\partial \over \partial \q_\alpha^a} x_{\alpha\dot\beta} \,,
\\
& \bar S_{b'\dot\beta} = x_{\alpha\dot\beta} \bq_{b'\dot\alpha} {\partial\over\partial x_{\alpha\dot\alpha}} +  y_{b'}{}^a 
{\partial\over\partial \theta_{\alpha}^a}  x_{\alpha\dot\beta}  - \bq_{a'\dot\beta} y_{b'}^a  {\partial\over\partial y_{a'}^a} + \bq_{b'\dot\alpha} \bq_{a'\dot\beta}
{\partial\over\partial \bq_{a' \dot\alpha}}\,.
\end{align} 
The explicit form of the $\mathcal N=4$ transformations generated by these operators can be found in Appendix \ref{App-trick}.

We notice that the super-propagator \re{D} is annihilated by all the generators \re{gens} except $\bar S_{\dot\alpha}^{a'}$. This allows us to simplify 
\re{G4-start} as
\begin{align}\label{G4-sim}
\mathcal G_4^{(\rm anom)} {}& =   Q^4 Q'{}^4  \bar S^4\widetilde S'{}^4 \left[ \theta_1^4\, \theta_2^4\, \theta_3^4\, \theta_4^4 \,  F(x)\right],
\end{align}
where the notation was introduced for 
\begin{align}\notag\label{tildeS}
\widetilde S_{b'\dot\beta} {}& = (D_{13} D_{24})^2 \bar S_{b'\dot\beta} (D_{13} D_{24})^{-2} %= \bar S_{\dot\alpha}^{a'} +2 \sum_{i=1}^4 \bar\theta_{i,\dot\alpha}^{a'}
\\
{}&  = \sum_i \bq_{i,b'\dot\alpha} {\partial\over\partial x_{i,\alpha\dot\alpha}}  x_{i,\alpha\dot\beta} +    y_{i,b'}{}^a 
{\partial\over\partial \theta_{i,\alpha}^a} x_{i,\alpha\dot\beta} - \bq_{i,a'\dot\beta} y_{i,b'}^a  {\partial\over\partial y_{i,a'}^a} + \bq_{i,b'\dot\alpha} \bq_{i,a'\dot\beta}
{\partial\over\partial \bq_{i,a' \dot\alpha}}  .
\end{align}
The only difference as compared with the last relation in \p{gens} is that, in the first term, $x_{\alpha\dot\beta}$ appears to the right of the derivative.  

As follows from \re{gens}, the generators $Q^\alpha_a$ and $\bar S_{a\dot\beta}$ do not involve spatial derivatives $\partial/\partial x$ 
and, therefore, in the expression on the right-hand side of \re{G4-sim}, they only act on the product of Grassmann variables. In this way, we arrive at 
the following relation \footnote{To evaluate $Q^4 \bar S^4 (\theta_1^4\dots \theta_4^4)$ it is convenient to use the integral
representation $Q^4 \theta_i = \int d^4 \epsilon \exp(Q\cdot\epsilon)\theta_i = \int d^4 \epsilon\, \theta_i' $ and similarly for $\bar S^4$ and apply relations \re{Q-tran}
and \re{sbar} from Appendix \ref{App-trick}.}
\begin{align}\label{master}
\mathcal G_4^{(\rm anom)}  &=  Q'{}^4\widetilde S'{}^4 \left[(x_{12}^2 x_{13}^2 x_{14}^2)^2  (x_{13}^{-1} \theta_{13} - x_{12}^{-1} \theta_{12})^4(x_{14}^{-1} \theta_{14} - x_{12}^{-1} \theta_{12})^4F(x)\right] ,
\end{align}
where the expression inside the  brackets is invariant under the exchange of any pair of points and 
is independent of $y_i$ and $\bar\theta_i $. This is the master formula that we shall use 
in the following section. 

According to \re{master}, the dependence of $\mathcal G_4^{(\rm anom)}$ on $y_i$ and $\bar\theta_i$ comes entirely from the product
of generators $Q'{}^4\widetilde S'{}^4=\prod_{a,\alpha} Q_{a'}^\alpha \prod_{b',\dot\beta}\widetilde S_{b'\dot\beta}$. Obviously, the expression on the right-hand side of \re{master} is not symmetric
under the exchange of chiral and antichiral sectors which seems to be in a contradiction with the expected reality property of the 
supercurrent $\mathcal T$ \cite{Howe:1996rb}. Nevertheless, this symmetry is restored owing to relation \re{I4}. If we use 
the second relation in \re{I4} we would derive another equivalent form of \re{master} in which the variables $\theta_i$ and $\bar\theta_i$
 are exchanged.
 
\section{Four-point correlation function of stress-energy tensors}\label{sect:TTTT}
 
In the previous section, we presented the general expression for the correlation function of the supercurrents, Eqs.~\re{G4}, 
\re{G4-rat}  and \re{master}. To obtain the four-point correlation function of the stress-energy tensors 
\begin{align}\label{G-TTTT}
G_{TTTT} (x) =  \vev{T_{\alpha_1\beta_1}^{\dot\alpha_1\dot\beta_1}(x_1)T_{\alpha_2\beta_2}^{\dot\alpha_2\dot\beta_2}(x_2)T_{\alpha_3\beta_3}^{\dot\alpha_3\dot\beta_3}(x_3)T_{\alpha_4\beta_4}^{\dot\alpha_4\dot\beta_4}(x_4)}
\end{align}
we have 
to go through two steps:  expand the supercorrelator $\mathcal G_4$ in powers of Grassmann variables, and then apply to $\mathcal G_4$ the 
four differential operators defined in \re{T-diff.op}, one for each stress-energy tensor. 

As follows from \re{G4}, the resulting expression for this correlation function can be decomposed into a sum of two terms, 
\begin{align}\label{T4-sum}
G_{TTTT} =  G_{TTTT}^{(0)}(x) +  G_{TTTT}^{(\rm anom)}(x)\,.
\end{align}
Here the first term $G_{TTTT}^{(0)}$ comes from \re{G4-rat}. It is a rational function of the distances $x_{ij}^2$, independent of the 
 variables $y_i$. Like the three-point correlation function, it can be decomposed into the sum of three different Lorentz
structures coinciding with the four-point correlation functions of stress-energy tensors in a free theory of scalars, fermions and gauge fields, respectively. To save space, we do not present their explicit expressions. 

Let us turn to the second term on the right-hand side of \re{T4-sum} generated by \re{master}. Since the differential operator
in \re{T-diff.op} is given by a sum of three terms, the operator that we have to apply to $\mathcal G_4^{(\rm anom)}$ in order
to extract $G_{TTTT}^{(\rm anom)}$ contains $3^4=81$ different terms. Nevertheless, as we show in this section, it is
 sufficient to examine only one term, the one containing the maximal number of derivatives
with respect to the Grassmann variables. It comes from the first term inside the brackets in \re{T-diff.op}
and selects the following component of the supercorrelator
\begin{align}\label{G4-aux}
\mathcal G_4^{(\rm anom)}  &= \prod_{i=1}^4 (\theta_i^2)^{\alpha_i\beta_i} (\bar\theta_i^2)_{\dot\alpha_i\dot\beta_i} \times G_{\alpha_1\beta_1\dots \alpha_4\beta_4}^{ \dot\alpha_1\dot\beta_1 \dots \dot\alpha_4\dot\beta_4} + \dots\,,
\end{align}
where $(\theta^2)^{\alpha\beta}= \theta^{\alpha a} \theta_{a}^{\beta}$, $(\bar\theta^2)_{\dot\alpha\dot\beta}= \bar\theta_{\dot\alpha a'} \bar\theta_{\dot\beta}^{a'}$ and dots denote the remaining components.

\subsection{Maximal number of derivatives recipe}\label{sect:trick}

Let us first compute $G_{\alpha_1\beta_1\dots \alpha_4\beta_4}^{ \dot\alpha_1\dot\beta_1 \dots \dot\alpha_4\dot\beta_4}$. Matching \re{G4-aux} to \re{master}, we observe
that the $8$ chiral variables $\theta_i$ on the right-hand side of \re{G4-aux} can only come from the expansion of the expression inside the 
brackets in \re{master}. Moreover, since the latter expression has the same degree $8$ in the $\theta$'s, we can neglect the 
terms containing derivatives $\partial_{\theta_i}$ in the product of generators $Q'{}^4\widetilde S'{}^4$ in \re{master}.
 In this way, from \re{gens} and \re{tildeS} we obtain a simplified form of the generators  $Q'$ and $\widetilde S'$, 
 \begin{align}\notag\label{sim-S}
{}&  Q^\alpha_{a'} = \sum_i \bq_{i,a'\dot\alpha} {\partial\over\partial x_{i,\alpha\dot\alpha}} \,,
\\
{}& \widetilde  { S} _{b'\dot\beta} =  \sum_i \bq_{i,b'\dot\alpha} {\partial\over\partial x_{i,\alpha\dot\alpha}}  x_{i,\alpha\dot\beta}  + \bq_{i,b'\dot\alpha} \bq_{i,a'\dot\beta}
{\partial\over\partial \bq_{i,a' \dot\alpha}}\,.
\end{align}    
Here in the second relation we also take into account that $(-\bq_{a'\dot\beta}\, y_{b'}^a \,  \partial_{y_{a'}^a})$ does not
contribute due to the $y-$independence of the expression inside the  brackets in \re{master}. 
 
In order to arrive at \re{G4-aux}, we have to extract the $(\bar\theta_1^2)_{\dot\alpha_1\dot\beta_1}\dots (\bar\theta_4^2)_{\dot\alpha_4\dot\beta_4}$ 
component from the product of generators $Q'{}^4\widetilde S'{}^4$ in \re{master}. This turns out to be an extremely nontrivial task,
mainly due to the presence of the second (nonlinear) term in the second line of \re{sim-S}. There is however a shortcut
that allows us to overcome this difficulty. In what follows we shall refer to it as the `maximal number of derivatives' recipe.

Let us examine the action of  $Q'{}^4\widetilde S'{}^4$ on a test function $f(x,\theta)$ independent of the variables $\bar\theta_i$.
Replacing the generators by their explicit expressions \re{sim-S} we find that the expansion of  $Q'{}^4\widetilde S'{}^4f(x,\theta)$ 
involves various terms including those containing the maximal number (equal to 8) of derivatives with respect to $x_i$.
To identify such terms we can safely neglect the second (nonlinear) term in $\widetilde  { S} _{b'\dot\beta}$ and, in addition, 
ignore the noncommutativity of $x_i$ and $\partial_{x_i}$. Introducing the notation $p_i=\partial_{x_i}$ and treating $x_i$ and $p_i$ as
commuting variables we get
\begin{align}\notag\label{trick}
 Q'{}^4\widetilde {S}^4f(x,\theta) {}&= \Big(\sum_{i=1}^4 \bq_i p_i\Big)^4  \Big( \sum_{i=1}^4 \bq_i p_i x_i \Big)^4f (x,\theta) + \dots
 \\
 {}&= {1\over (x_{34}^2)^2} 
 \Big(\sum_{i} \bq_i p_i x_{i3}\Big)^4 \Big(\sum_{i} \bq_i p_i x_{i4}\Big)^4 f (x,\theta) + \dots\,,
\end{align}
where we used the shorthand notation $(\bq_i p_i)^\alpha_{a'}\equiv  \bq_{i,a'\dot\alpha} p_i^{\dot\alpha\alpha}$ and $(\bq_i p_i x_i)_{b'\dot\beta} \equiv \bq_{i,b'\dot\alpha}   p_i^{\dot\alpha\alpha}  x_{i,\alpha\dot\beta}$ and denoted by dots terms with fewer spatial derivatives. By construction, \re{trick} only describes the terms containing $8$ spatial derivatives acting on a test
function.  

We are now ready to formulate the maximal number of derivatives recipe. It consists of two steps: (i) expand \p{trick} in powers of 
$\bq_i$ and move {all} $p_i$ to the left of $x_i$, and then (ii) replace $p_i=\partial_{x_i}$ so that the derivatives act on {all} 
$x_i$ to the right. It turns out that the resulting expression gives the {\em exact} result for $Q'{}^4\widetilde {S}^4f(x,\theta)$, including
the terms shown by dots in \re{trick}. The proof of this statement can be found in Appendix \ref{App-trick}.
 
Let us now apply the above  recipe to identify the $(\bar\theta_1^2)_{\dot\alpha_1\dot\beta_1}\dots (\bar\theta_4^2)_{\dot\alpha_4\dot\beta_4}$  component of $Q'{}^4\widetilde {S}^4f(x,\theta)$. We first use the second relation in \re{trick} to get 
\begin{align}\label{M-ind}
Q'{}^4\widetilde {S}^4f(x,\theta) =\bigg( {\prod_{i=1}^4 (\bar\theta_i^2)_{\dot\alpha_i\dot\beta_i}
p_i^{\dot\alpha_i\alpha_i}p_i^{\dot\beta_i\beta_i} }\bigg)
\mathcal M_{\alpha_1\beta_1\dots \alpha_4\beta_4}(x) f(x,\theta) + \dots\,,
\end{align}
where $\mathcal M_{\alpha_1\beta_1\dots \alpha_4\beta_4}(x)$ is   symmetric under the exchange of the spinor indices $\alpha_i\leftrightarrow \beta_i$. It admits
a concise representation in terms of the auxiliary spinors introduced in \re{T-pro}
\begin{align}\label{cal-M}
\mathcal M_{\alpha_1\beta_1\dots \alpha_4\beta_4}(x) = {\partial^2 \over \partial  \lambda_{1}^{\alpha_1}\partial  \lambda_{1}^{ \beta_1}}\dots {\partial^2 \over \partial \lambda_{4}^{\alpha_4}\partial  \lambda_{4}^{\beta_4}} \mathcal M(x,\lambda)\,,
\end{align} 
where $\mathcal M(x,\lambda)$ is given by 
\begin{align}\label{M4}
\mathcal M(x,\lambda)= \Big(\vev{1|x^{-1}_{12}x_{24}|4} \vev{3|x_{31}x_{12}|2}  - \vev{1|x^{-1}_{12}x_{23}|3} \vev{4|x_{41}x_{12}|2}\Big)^2\,,
\end{align}
with $\vev{1|x^{-1}_{12}x_{24}|4}= \lambda_1^\alpha (x^{-1}_{12}x_{24})_\alpha{}^\beta \lambda_{4,\beta}$.
Obviously, the expression on the right-hand side of \re{cal-M} does not depend on the auxiliary spinors  $\lambda_i$. They were introduced
to make the symmetry of \re{cal-M} under the exchange of indices more transparent.
It is straightforward to verify that $\mathcal M(x,\lambda)$ is invariant under the exchange of any pair of points, $(x_i,\lambda_i)\leftrightarrow (x_j,\lambda_j)$.  In addition, applying \re{inv-la} with $w(x_i)=1$ we find that it transforms covariantly under inversion  
\begin{align}\label{inv-M4}
I\big[ \mathcal M(x,\lambda) \big] =   (x_1^2 x_2^2 x_3^2 x_4^2)^{-2} \mathcal M(x,\lambda)\,.
\end{align}

Let us now go through the second step of the recipe. Since all $p_i$ in \re{M-ind} are located to the left of some $x-$dependent functions,  we replace $p_i=\partial_{x_i}$ on the right-hand side of \re{M-ind} and obtain the exact expression for this particular component of $Q'{}^4\widetilde {S}^4f(x,\theta)$. Notice that in the resulting expression the spatial derivatives act both on $\mathcal M_{\alpha_1\beta_1\dots \alpha_4\beta_4}(x)$ and on the test function $f(x,\theta)$.

Going back to  \re{master} we choose the test function in \re{M-ind} to match the expression inside
the  brackets in  \re{master},
\begin{align}
f(x,\theta)=(x_{12}^2 x_{13}^2 x_{14}^2)^2  (x_{13}^{-1} \theta_{13} - x_{12}^{-1} \theta_{12})^4(x_{14}^{-1} \theta_{14} - x_{12}^{-1} \theta_{12})^4F(x)\,.
\end{align}
In order to compute \re{G4-aux} we only need its component proportional to $(\theta_1^2)_{\alpha_1\beta_1} \dots (\theta_4^2)_{\alpha_4\beta_4}$.
Going through the calculation we find
\begin{align}
f(x,\theta)=   \prod_i (\theta_i^2)^{\alpha_i\beta_i}\mathcal M_{\alpha_1\beta_1\dots \alpha_4\beta_4}(x)  F(x) +\dots\,,
\end{align}
where $\mathcal M_{\alpha_1\beta_1\dots \alpha_4\beta_4}(x)$ is given by \re{cal-M} and the dots denote other components.  
Finally, we substitute the last relation into \re{M-ind} and match the result to \re{G4-aux} to find
\begin{align}\label{G4-der}
G_{\alpha_1\beta_1\dots \alpha_4\beta_4}^{ \dot\alpha_1\dot\beta_1 \dots \dot\alpha_4\dot\beta_4}= {\prod_{i=1}^4  
(\partial_{x_i})^{\dot\alpha_i\gamma_i}(\partial_{x_i})^{\dot\beta_i\delta_i} }\Big[\mathcal M_{\alpha_1\beta_1\dots \alpha_4\beta_4}(x) 
\mathcal M_{\gamma_1\delta_1\dots \gamma_4\delta_4}(x) F(x)\Big]\,.
\end{align} 
Here the nontrivial information about the particular $\mathcal N=4$ superconformal theory is encoded in the  
function $F(x)$ defined in \re{F-Phi}.

\subsection{Restoration of the symmetry}

We recall that relation \re{G4-der} defines a particular component of the supercorrelator \re{G4-aux}. To get the four-point
correlation function of the stress-energy tensors \re{G-TTTT}, we have to add to \re{G4-der} the contribution of the remaining $80$ terms 
mentioned in the beginning of this section. To understand their role, let us check whether  \re{G4-der} is consistent with the expected 
properties of the stress-energy tensor.

To simplify the analysis, consider the correlation function 
\begin{align}
G_{\alpha_1\beta_1}^{ \dot\alpha_1\dot\beta_1} = \vev{T_{\alpha_1\beta_1}^{ \dot\alpha_1\dot\beta_1}(x_1) \dots }\,,
\end{align}
where the dots denote other operators. Then, the symmetry and conservation properties of the stress-energy tensor, $T_{\mu\nu}-T_{\nu\mu}=g^{\mu\nu}T_{\mu\nu}=
\partial^\mu T_{\mu\nu}=0$, yield the Ward identities \footnote{Strictly speaking, relation \re{WI} is valid up to contact terms \cite{Fradkin:1978pp,Osborn:1993cr,Erdmenger:1996yc}. For four-point correlation functions in an $\mathcal N=4$ superconformal theory, the contact terms are only due to the rational part of \re{T4-sum}, whereas the anomalous contribution to \re{T4-sum} is less singular and satisfies homogenous Ward 
identities. }
\begin{align}\label{WI}
G_{\alpha_1\beta_1}^{ \dot\alpha_1\dot\beta_1} - G_{\beta_1\alpha_1}^{\dot\beta_1\dot\alpha_1} = \epsilon^{\alpha_1\beta_1}\epsilon_{\dot\alpha_1\dot\beta_1} G_{\alpha_1\beta_1}^{\dot\alpha_1\dot\beta_1}  =  (\partial_{x_1})^{\alpha_1}_{\dot\alpha_1} G_{\alpha_1\beta_1}^{ \dot\alpha_1\dot\beta_1} =0\,.
\end{align}
In addition, $G_{\alpha_1\beta_1}^{\dot\alpha_1\dot\beta_1}$ should transform covariantly under conformal transformations with weight at point $x_1$ corresponding to a conformal primary operator of Lorentz spin $(1,1)$ and dimension $4$.

Let us verify whether \re{G4-der} satisfies relations \re{WI}. It is easy to see that the first two relations in \re{WI} are automatically
satisfied due to the symmetry of \re{cal-M} under the exchange of indices, $\epsilon^{\alpha_1\beta_1} \mathcal M_{\alpha_1\beta_1\dots \alpha_4\beta_4}(x)=0$. Examining the last relation in \re{WI} we find, using the identity $\partial_{\dot\alpha}^\alpha \partial ^{\dot\alpha\gamma}= \epsilon^{\alpha\gamma}\Box$,  
\begin{align}\label{non0}
(\partial_{x_1})^{\alpha_1}_{\dot\alpha_1} G_{\alpha_1\beta_1\dots \alpha_4\beta_4}^{ \dot\alpha_1\dot\beta_1 \dots \dot\alpha_4\dot\beta_4}\sim \epsilon^{\alpha_1\gamma_1} \Box_{x_1}\Big[\mathcal M_{\alpha_1\beta_1\dots \alpha_4\beta_4}(x) 
\mathcal M_{\gamma_1\delta_1\dots \gamma_4\delta_4}(x) F(x)\Big].
\end{align}
Since this expression is different from zero for a generic function $F(x)$, we conclude that \re{G4-der} does not respect the conservation
of the stress-energy tensor. This is not surprising since \re{G4-der} is only a part of the correlation function \re{G-TTTT} and
for the expression on the right-hand side of \re{non0} to vanish we have to add the contributions of the remaining 80 terms.

We notice that \re{non0} would automatically vanish if the expression inside the  brackets on the right-hand side of \re{non0} 
and \re{G4-der} are symmetric under the exchange of indices $\alpha_1$ and $\gamma_1$. Similarly, the vanishing
of $(\partial_{x_i})^{\alpha_i}_{\dot\alpha_i} G_{\alpha_1\beta_1\dots \alpha_4\beta_4}^{ \dot\alpha_1\dot\beta_1 \dots \dot\alpha_4\dot\beta_4}$ would follow from the symmetry under $\alpha_i\leftrightarrow \gamma_i$. This suggests that the net effect of
the remaining $80$ terms is to symmetrize the product of two $\mathcal M-$tensors in \re{G4-der} with respect to the 
indices $(\alpha_i,\beta_i,\gamma_i,\delta_i)$. Taking into account \re{cal-M} we find that {(up to an overall normalization factor)} this amounts to replacing in \re{G4-der}
\begin{align}\label{subs}
\mathcal M_{\alpha_1\beta_1\dots \alpha_4\beta_4}(x) 
\mathcal M_{\gamma_1\delta_1\dots \gamma_4\delta_4}(x) \  \ \longrightarrow \ \ \prod_{i}
(\partial_{\lambda_i})_{\alpha_i}(\partial_{\lambda_i})_{\beta_i}(\partial_{\lambda_i})_{\gamma_i}(\partial_{\lambda_i})_{\delta_i}\left[\mathcal M(x,\lambda)\right]^2,
\end{align}
with $\mathcal M(x,\lambda)$ given by \re{M4}. Upon this substitution, \re{G4-der} respects the conservation of the stress-energy tensor.
    
Let us now examine the conformal properties of \re{G4-der}.  As before, it is convenient to examine the inversion. 
Denoting the expression inside the  brackets in \re{G4-der} by $F_{\alpha\beta\gamma\delta}$, we find using \re{inv-F}, \re{cal-M}
and \re{inv-M4} that it transforms covariantly
\begin{align}
I\left[ F_{\alpha\beta\gamma\delta}\right] = \prod_i (x_i)^{\alpha_i'}_{\dot\alpha_i}(x_i)^{\beta_i'}_{\dot\beta_i}(x_i)^{\gamma_i'}_{\dot\gamma_i}(x_i)^{\delta_i'}_{\dot\delta_i}\, F_{\alpha'\beta'\gamma'\delta'}\,.
\end{align}
Then, we use $G_{\alpha_1\beta_1\dots \alpha_4\beta_4}^{ \dot\alpha_1\dot\beta_1 \dots \dot\alpha_4\dot\beta_4}= {\prod_{i=1}^4  
\partial_{x_i}^{\dot\alpha_i\gamma_i}\partial_{x_i}^{\dot\beta_i\delta_i} }F_{\alpha\beta\gamma\delta}$ and apply the identity \re{invs} to find  
\begin{align}\label{nonhom}\notag
I[G_{\alpha_1\beta_1\dots \alpha_4\beta_4}^{ \dot\alpha_1\dot\beta_1 \dots \dot\alpha_4\dot\beta_4}]
{}& = \prod_i  (x_i^2)^2 
(x_i)^{\alpha_i}_{\dot\gamma_i'} \, (x_i)^{\beta_i}_{\dot\delta_i'}  \, \partial_{x_i}^{\dot\gamma_i'\gamma_i'}\partial_{x_i}^{\dot\delta_i'\delta_i'}   (x_i)^{\alpha_i'}_{\dot\alpha_i}(x_i)^{\beta_i'}_{\dot\beta_i} \,
 F_{\alpha'\beta'\gamma'\delta'}
 \\
{}& =\prod_i  (x_i^2)^2 
(x_i)^{\alpha_i}_{\dot\gamma_i'} \, (x_i)^{\beta_i}_{\dot\delta_i'}  \,  (x_i)^{\alpha_i'}_{\dot\alpha_i}(x_i)^{\beta_i'}_{\dot\beta_i} \,
G_{\alpha_1'\beta_1'\dots \alpha_4'\beta_4'}^{ \dot\gamma_1'\dot\delta_1' \dots  \dot\gamma_4'\dot\delta_4'} + \dots
 \,,
\end{align}
where the dots in the second relation denote inhomogenous `bad' terms coming from the commutators $[\partial_{x_i}^{\dot\gamma_i'\gamma_i'}\partial_{x_i}^{\dot\delta_i'\delta_i'} ,  x^{\alpha_i'}_{i,\dot\alpha_i}x^{\beta_i'}_{i,\dot\beta_i}]$. Due to the presence of such terms, 
$G_{\alpha_1\beta_1\dots \alpha_4\beta_4}^{ \dot\alpha_1\dot\beta_1 \dots \dot\alpha_4\dot\beta_4}$ does not transform covariantly. Their contribution however 
involves terms like $\epsilon^{\alpha'\gamma'}F_{\alpha'\beta'\gamma'\delta'}$ and it would vanish if $F_{\alpha'\beta'\gamma'\delta'}$ is symmetric with respect to the 
indices $\alpha'$ and $\gamma'$. This is exactly what happens if we apply \re{subs} to \re{G4-der}. 

Thus, we conclude that the substitution \re{subs} not only ensures the conservation of the stress-energy tensor but also 
restores the correct conformal properties of the four-point correlation function.
 
The question remains however whether \re{subs} correctly describes the additional contribution to \re{G4-der} 
coming from the last two terms on the right-hand side of \re{T-diff.op}. Going through a lengthy and tedious calculation we
verified that this is indeed the case. 

\subsection{Simplified form of the correlation function}

Combining together \re{G4-der} and \re{subs}, we arrive at the following remarkably simple result for the four-point correlation
function of stress-energy tensors
\begin{align}\label{TTTT-ini}
G_{\alpha_1\beta_1\dots \alpha_4\beta_4}^{ \dot\alpha_1\dot\beta_1 \dots \dot\alpha_4\dot\beta_4}(x)= {\prod_{i=1}^4  
(\partial_{x_i})^{\dot\alpha_i\gamma_i}(\partial_{x_i})^{\dot\beta_i\delta_i} }
(\partial_{\lambda_i})_{\alpha_i}(\partial_{\lambda_i})_{\beta_i}(\partial_{\lambda_i})_{\gamma_i}(\partial_{\lambda_i})_{\delta_i}\Big\{ \! \left[\mathcal M (x,\lambda)\right]^2 
  F(x)\Big\}\,.
\end{align}
More precisely, this relation describes the `anomalous' (non-rational) contribution to \re{T4-sum}.   
We recall that  $F(x)$ is the nontrivial dynamical function defined 
in \re{F-Phi} and $\mathcal M (x,\lambda)$ is a kinematical function  given by \re{M4}. Relation \re{TTTT-ini} is one of the main results of this paper.

Since $\mathcal M (x,\lambda)$ is a homogenous polynomial  of degree $2$  in $\lambda_i^\alpha$,
the expression on the right-hand side \re{TTTT-ini} does not depend on the auxiliary spinors $\lambda_i$. This polynomial
admits an elegant representation when expressed in terms of new variables 
\begin{align}\label{Z}
Z_i^I = \left(\lambda_i^\alpha \atop (x_i)^{\dot\alpha\beta} \lambda_{i\beta}\right) \equiv  \left(\ket{i}  \atop  x_i \ket{i}\right)
\,,\qquad \text{ $I=(\alpha,\dot\alpha)$},
\end{align}
which satisfy the following determinant relation  
\begin{align} \notag\label{1234}
\vev{1\, 2\, 3\,4} {}& \equiv \epsilon_{IJKL} Z_1^I Z_2^J Z_3^K Z_4^L
\\[2mm]
{}& =\vev{1|x^{-1}_{12}x_{23}|3} \vev{4|x_{41}x_{12}|2}-\vev{1|x^{-1}_{12}x_{24}|4} \vev{3|x_{31}x_{12}|2}   \,.
\end{align}
Then, we find from  \re{M4}
\begin{align}
\mathcal M (x,\lambda) = \vev{1\, 2\, 3\,4}^2\,.
\end{align}
In this representation, the properties of $\mathcal M (x,\lambda)$ becomes more transparent.

Relations \re{Z} and \re{1234} are very familiar from the dual space description of scattering amplitudes.
There, the variables $\lambda_i$ and $p_i=x_i-x_{i+1}$ define the helicity and the {on-shell} momentum of each scattered particle, 
respectively, and the dual conformal symmetry is realized as $SL(4)$ transformations of $Z_i^I$. Relation 
\re{1234} defines the simplest four-point dual conformal invariant and serves as a building block in constructing 
four-dimensional integrands for the scattering amplitudes  in planar $\mathcal N=4$ SYM \cite{Mason:2009qx,ArkaniHamed:2010kv}.  

As explained in the previous subsection,  relation \re{TTTT-ini} respects the conservation and conformal symmetry of the stress-energy tensor, independently
of the form of $F(x)$. The expression on the right-hand side of \re{TTTT-ini} involves $8$ spatial derivatives acting on the product of two functions. Obviously, its 
expansion yields a very lengthy expression involving the 22 different Lorentz structures mentioned in the Introduction, each of them involving derivatives of
$F(x)$. The very fact that the correlation function in an $\mathcal N=4$ superconformal theory admits a compact representation \re{TTTT-ini} leads to important 
consequences that we explain in section \ref{sect:app}.

We can further simplify \re{TTTT-ini} by considering the correlation function of the operators \re{T-pro}. This amounts to 
projecting the Lorentz indices on both sides of \re{TTTT-ini} with the auxiliary spinors $\lambda_i$ and $\bar\lambda_i$  
\begin{align}
G_{TTTT}^{(\rm anom)} {}& = {1\over 4^4}\prod_{i=1}^4   \lambda_i^{\alpha_i} \lambda_i^{\beta_i}\bar\lambda_{i,\dot\alpha_i} \bar\lambda_{i,\dot\beta_i}
\, G_{\alpha_1\beta_1\dots \alpha_4\beta_4}^{ \dot\alpha_1\dot\beta_1 \dots \dot\alpha_4\dot\beta_4}\,.
\end{align}
Introducing the notation for the differential operator
\begin{align}\label{Di}
\mathcal D_i = \bar\lambda_{i,\dot\alpha} (\partial_{x_i})^{\dot\alpha\alpha}(\partial_{\lambda_i})_{\alpha}\equiv [i| \partial_{x_i} \ket{\partial_{\lambda_i}}\,,
\end{align}
we finally obtain from \re{TTTT-ini} a very compact representation for the correlation function 
\begin{align}\label{TTTT-com}
G_{TTTT}^{(\rm anom)} = 4^4 \,(\mathcal D_1  \mathcal D_2\mathcal D_3\mathcal D_4)^2\Big[ \vev{1\, 2\, 3\,4}^4
  F(x)\Big].
\end{align}
Here the additional normalization factor comes from $\lambda_i^{\alpha}(\partial_{\lambda_i})_{\alpha} \vev{1\, 2\, 3\,4}^4 = 4 \,\vev{1\, 2\, 3\,4}^4$.

We notice that, at first glance,  relations \re{TTTT-ini} and \re{TTTT-com} are not symmetric under the exchange of the chiral and antichiral sectors,  
 $\alpha_i \leftrightarrow \dot\alpha_i$ and $\beta_i \leftrightarrow \dot\beta_i$. As was already mentioned in the previous section, thanks to  relation \re{I4},
 the correlation functions \re{TTTT-ini} and \re{TTTT-com}  admit another equivalent representation in which the chiral and antichiral indices are interchanged. 

\section{Four-point correlation functions of currents}\label{sect:curr}

In the previous section we demonstrated that the calculation of the correlation function of stress-energy tensors can be greatly simplified
by employing the maximal number of derivatives recipe. It allowed us to obtain the very special representation \re{G4-der} for the relevant component 
of the supercorrelator $\mathcal G_4$, and then to promote it to the complete expression for the correlation function by simply symmetrizing the product of Lorentz
tensors with respect to the chiral indices, Eq.~\re{subs}. 

In this section we argue that the same approach can be applied to the four-point correlation functions involving other components 
of the stress-energy supermultiplet, for example,  the $R-$symmetry current \re{J}.

\subsection{Ansatz}\label{sect:ans}

Instead of going through the computation of the various correlation functions, let us try to generalize \re{TTTT-ini} and formulate  an ansatz
for the correlation function involving conserved currents $J_{\alpha_1 \dots \alpha_S}^{\dot\alpha_1\dots  \dot\alpha_S}(x)$ of Lorentz spin $(S/2,S/2)$ 
\begin{align}\label{GS}
G_{\alpha_1 \dots \alpha_S}^{\dot\alpha_1\dots  \dot\alpha_S}(x,\dots ) = \vev{0|J_{\alpha_1 \dots \alpha_S}^{\dot\alpha_1\dots  \dot\alpha_S}(x) \dots |0}\,,
\end{align}
where the dots denote the remaining operators.  For $S=1$ and $S=2$ the operator $J_{\alpha_1 \dots \alpha_S}^{\dot\alpha_1\dots  \dot\alpha_S}(x)$
coincides with the $R-$symmetry current and the stress-energy tensor, respectively. In close analogy with \re{G4-der}, we assume that 
the correlation function \re{GS} has the following general form~\cite{Belitsky:2014zha}
\begin{align}\label{spin-S}
G_{\alpha_1 \dots \alpha_S}^{\dot\alpha_1\dots  \dot\alpha_S}(x,\dots ) = \partial_x^{\dot\alpha_1\gamma_1} \dots  \partial_x^{\dot\alpha_S \gamma_S} \Big[
\mathcal M_{\alpha_1\gamma_1\dots \alpha_S\gamma_S}(x)F(x)\Big]\,,
\end{align}
where $ \partial_x^{\dot\alpha  \gamma }\equiv \partial/\partial {x_{\gamma \dot\alpha }}$ and the number of spatial derivatives matches the Lorentz spin
of the operator. If other operators in the correlation function \re{GS} carry nonvanishing Lorentz spin, the expression on the right-hand side of \re{spin-S} 
contains additional derivatives acting on their coordinates. 
 
To ensure the conservation of the current, we require that  $\mathcal M_{\alpha_1\gamma_1\dots \alpha_S\gamma_S}(x,\dots)$ be completely symmetric with respect to all $2S$ chiral
indices. Indeed, we verify that
\begin{align}
(\partial_x)_{\dot\alpha_1}^{\alpha_1}G_{\alpha_1 \dots \alpha_S}^{\dot\alpha_1\dots  \dot\alpha_S}(x,\dots ) \sim   \epsilon^{\alpha_1\gamma_1} \Box_x\,\Big[
\mathcal M_{\alpha_1\gamma_1\dots \alpha_S\gamma_S}(x)F(x)\Big]= 0
\end{align}
for an arbitrary $x-$dependent completely symmetric tensor $\mathcal M_{\alpha_1\gamma_1\dots \alpha_S\gamma_S}(x,\dots)$. Let us also demand  that 
\re{spin-S} has the correct transformation properties under inversion
\footnote{We recall that inversion swaps the chirality of the Lorentz indices.}
\begin{align}\label{Inv-S}
I[G_{\alpha_1 \dots \alpha_S}^{\dot\alpha_1\dots  \dot\alpha_S}(x,\dots ) ] {}& = (x^2)^2 
 x^{\alpha_1}_{\dot\beta_1}  x_{ \dot\alpha_1}^{\beta_1}\ldots x^{\alpha_S}_{\dot\beta_S}  x_{ \dot\alpha_S}^{\beta_S} G_{\beta_1 \dots \beta_S}^{\dot\beta_1\dots  \dot\beta_S}(x,\dots ) \,,
\end{align}
where we do not display the weight factors corresponding to the other operators. Taking into account \re{inv-F} and the identity \re{ten-inversion}, we find that \re{Inv-S} and \re{spin-S} 
lead to  
\begin{align}\label{Inv-M}
I\left[\mathcal M_{\alpha_1\gamma_1\dots \alpha_S\gamma_S}(x)
 \right] =(x^2)^{-S-2}    x^{\beta_1}_{\dot\alpha_1}x^{\delta_1}_{\dot\gamma_1} \dots  x^{\beta_S}_{\dot\alpha_S}x^{\delta_S}_{\dot\gamma_S}\,
 \mathcal M_{\beta_1\delta_1\dots \beta_S\delta_S}(x) \,.
\end{align}
As before, we can simplify this relation by projecting all chiral Lorentz indices by the auxiliary spinor $\lambda$,
\begin{align}
\mathcal M_S(x,\lambda)=\lambda^{\alpha_1} \lambda^{\gamma_1}  \dots \lambda^{\alpha_S} \lambda^{\gamma_S}  \mathcal M_{\alpha_1\gamma_1\dots \alpha_S\gamma_S}(x)\,.
\end{align}
Taking into account the transformation properties of the auxiliary spinors under inversion \re{inv-la}, we find from \re{Inv-M}
\begin{align}
I[ \mathcal M_S(x,\lambda)] = w^{2S}(x)  (x^2)^{-S-2}\mathcal M_S(x,\lambda)\,,
\end{align}
where $w(x)=(x^2)^n$ with an arbitrary $n$.  In the previous section we used $w=1$ but it is now more convenient to choose $w=x^2$, 
so that $I[\lambda^\alpha] =  x^{\dot\alpha\beta} \lambda_{\beta}$. 

The $x-$dependent factor on the right-hand side of the last relation defines the local conformal weight associated with 
the current of spin $S$. If the correlation function \re{GS} contains four currents with different spins $S_i$, the total 
conformal weight of the corresponding function $\mathcal M_{S_1 S_2 S_3 S_4}(x,\lambda)$ is given by the product
of four such factors, one for each current. As a consequence, $\mathcal M_{S_1 S_2 S_3 S_4}(x,\lambda)$  depends on four auxiliary spinors $\lambda_i$ (with $i=1,\dots,4$) and satisfies two 
main requirements: (i) to be a homogenous polynomial in $\lambda_i^\alpha$ of degree $2S_i$ and (ii) to transform under inversion as
 \begin{align}\label{M-eq}
I[ \mathcal M_{S_1 S_2 S_3 S_4}(x,\lambda)] =\prod_i   (x_i^2)^{S_i-2} \,\mathcal M_{S_1 S_2 S_3 S_4}(x,\lambda)\,.
\end{align}
In addition, if two operators are identical with, e.g., $S_1=S_2$,  in virtue of the Bose symmetry of the correlation function, the function 
$\mathcal M_{S_1 S_1 S_3 S_4}(x,\lambda)$ should be invariant under the exchange of points $1$ and $2$. 

Having determined $\mathcal M_{S_1 S_1 S_3 S_4}(x,\lambda)$, we can reconstruct the four-point correlation function of
the currents $J_{\alpha_1 \dots \alpha_{S_i}}^{\dot\alpha_1\dots  \dot\alpha_{S_i}}(x_i)$. To simplify the resulting expression, we project 
all Lorentz indices by the  (anti)chiral auxiliary spinors $\lambda_i$ and $\bar\lambda_i$ and define
\begin{align}
G_{S_1 S_2 S_3 S_4}(x,\lambda,\bar\lambda) {}& = \lambda_1^{\alpha_1}\dots \lambda_4^{\delta_{S_4}} \bar\lambda_{1,\dot\alpha_1}\dots \bar\lambda_{4,\dot\delta_{S_4}}
G_{\alpha_1 \dots \alpha_{S_1},\beta_1 \dots \beta_{S_2},\gamma_1 \dots \gamma_{S_3},\delta_1 \dots \delta_{S_4}}^{\dot\alpha_1\dots  \dot\alpha_{S_1},\dot\beta_1\dots  \dot\beta_{S_2},\dot\gamma_1 \dots \dot\gamma_{S_3},\dot\delta_1 \dots \dot\delta_{S_4}}(x)\,.
\end{align}
This function admits the following representation (up to an overall normalisation factor)
\begin{align} \label{G-spins}
G_{S_1 S_2 S_3 S_4}(x,\lambda,\bar\lambda)=  \mathcal D_1^{S_1}  \mathcal D_2^{S_2} \mathcal D_3^{S_3}\mathcal D_4^{S_4}\Big[ \mathcal M_{S_1 S_2 S_3 S_4}(x,\lambda)  F(x)\Big],
\end{align}
where the differential operator $\mathcal D_i$ is defined in \re{Di}. This relation generalizes \re{TTTT-com} to the case of currents 
of  spin $S_i=1,2$. Moreover, it is also applicable for $S_i=0$ in which case the corresponding operator is the $1/2$ BPS 
operator $O_{\bf 20'}$.
Notice that the $R-$symmetry current and the $1/2$ BPS operator have an $R-$charge and depend on the auxiliary variables $y$. In the 
representation \re{G-spins}, this dependence is carried by the function $\mathcal M_{S_1 S_2 S_3 S_4}(x,\lambda)$.
 
As shown in \cite{Belitsky:2014zha},  relations \re{spin-S} and \re{G-spins} can be generalized to the fermionic ${\cal N}=4$ supersymmetry currents 
$J_{\alpha_1\alpha_2}^{\dot\alpha}$ and $J^{\dot\alpha_1\dot\alpha_2}_{\alpha}$, which are also  members of the stress-energy supermultiplet.
They carry Lorentz spins  $(3/2,1/2)$ and $(1/2,3/2)$, respectively, and should appear in pairs in a non-vanishing correlation function. The only difference as compared with \re{spin-S} is that the numbers of (chiral) $\alpha-$ and 
$\gamma-$indices of the fully symmetric tensor $\mathcal M$ do not match. As a result, in the general
case of operators of Lorentz spin $(S_i/2, S_i'/2)$, their four-point correlation function is given by \re{G-spins} with
$\mathcal D_i^{S_i}$ replaced by $\mathcal D_i^{S_i'}$. As already mentioned, there exists another, equivalent representation 
of the same correlation function involving the conjugate operators $\bar D_i^{S_i}=[\partial_{\bar\lambda_i}| \partial_{x_i} \ket{i}^{S_i}$ and conjugate antichiral tensor $ {\mathcal {\widebar  M}}_{S_1'S_2'S_3'S_4'}(x,\bar \lambda)$.
 
\subsection{Special solutions} 

Let us construct some solutions to \re{M-eq}. We notice that \re{M-eq} defines $\mathcal M_{S_1 S_2 S_3 S_4}(x,\lambda)$ 
up to multiplication by an arbitrary function of the conformal cross-ratios $u$ and $v$ defined in \re{F-Phi}. Such a function will in general induce additional singularities for $x_{ij}^2\to 0$. To fix the ambiguity, we shall assume that $\mathcal M_{S_1 S_2 S_3 S_4}(x,\lambda)$ should not have such singularities.

We expect that $\mathcal M_{S_1 S_2 S_3 S_4}(x,\lambda)$ is a rational function of the distances $x_{ij}^2$
admitting an analytic continuation to complex space-time coordinates. The rational behind this is that the analysis 
of the conformal properties of $\mathcal M_{S_1 S_2 S_3 S_4}(x,\lambda)$ can be simplified by employing Dirac's embedding
formalism. There the complexified Minkowski space is realized as a light-cone in complex projective space $\mathbb{CP}^5$ with homogenous coordinates $X^{IJ}=-X^{JI}$ (with $I,J=1,\dots,4$) satisfying $\epsilon_{IJKL} X^{IJ} X^{KL}=0$.
The complex coordinates $x^{\dot\alpha\alpha}$ define a particular parameterization of $X^{IJ}$
\begin{align}\label{X-proj}
X^{IJ} = \left[\begin{array}{cl} \epsilon^{\alpha\beta} & -x^{\dot\beta\alpha}  \\ x^{\dot\alpha\beta} &  x^2 \epsilon^{\dot\alpha\dot\beta} \end{array}\right], \qqqquad X_{IJ}=\frac12 \epsilon_{IJKL} X^{KL}= \left[\begin{array}{cc}  x^2\epsilon_{\alpha\beta} & -x_{\alpha\dot\beta}  \\ x_{\beta\dot\alpha} &   \epsilon_{\dot\alpha\dot\beta} \end{array}\right]\,,
\end{align}
with composite indices $I=(\alpha,\dot\alpha)$ and $J=(\beta,\dot\beta)$. Then, the conformal transformations correspond to 
global $SL(4;\mathbb{C})$ transformations of $X^{IJ}$. The attentive reader will notice the similarity between \re{X-proj} and
\re{Y-proj}. Indeed, $x_{\alpha\dot\alpha}$ and $y_{aa'}$ appear on an equal footing as bosonic coordinates of the  supercurrent.

In addition to \re{X-proj}, we also need the variables \re{Z} that carry the dependence on the chiral spinors $\lambda$.
As was already mentioned, there exists some freedom in choosing the weight factor $w$ in \re{inv-la}.
The advantage of the choice $w=x^2$ is that the action of inversion on $Z^I$ corresponds to a global $SL(4)$ transformation
\footnote{In fact, $Z^I$ can be identified as coordinates on twistor space.}
\begin{align}
I[Z_i^I]  = \left(x_i\ket{\lambda_i} \atop -\ket{\lambda_i}\right) = \Omega^I_J \, Z_i^J\,.
\end{align}

We can use the variables $X_i^{IJ}$ and $Z_i^I$ (with $i=1,\dots,4$) to define various $SL(4)$ invariant quantities. 
Taking into account the identity $Z_i^I (X_i)_{IJ} = 0$ we can define three different $Z-$dependent structures
\begin{align}\notag\label{Xi}
{}& \vev{1\,2\,3\,4}= \epsilon_{IJKL} Z_1^I Z_2^J Z_3^K Z_4^L  \,,
\\[2mm] \notag
{}& \mathcal X_{[12]3} = Z_1^I (X_3)_{IJ} Z_2^J \sim \vev{1|x_{13} x_{32}|2}\,,
\\[2mm]
{}&\mathcal X_{1[234]} =  Z_1^I (X_2)_{IJ} (X_3)^{JK} (X_4)_{KL}(Z_1)^L \sim \vev{1|x_{12} x_{23}x_{34} x_{41}|1}\,,
\end{align}
where the first structure already appeared in \re{1234}.   $\mathcal X_{[12]3}$ and $\mathcal X_{1[234]}$
are antisymmetric with respect to the points indicated inside the  brackets, whereas $ \vev{1\,2\,3\,4}$ is completely
antisymmetric with respect to the four points. We verify that, as expected, they transform covariantly under inversion
\begin{align}\label{basis}
I[\vev{1\,2\,3\,4}] = \vev{1\,2\,3\,4}\,,\qquad
I[\mathcal X_{[12]3}] ={1\over x_3^2} \mathcal X_{[12]3}  \,,\qquad
I[\mathcal X_{1[234]}] ={1\over x_2^2x_3^2x_4^2} \mathcal X_{1[234]}\,.
\end{align}
We are now ready to construct solutions to \re{M-eq}.

Let us first revisit the four-point correlation function of stress-energy tensors,  $S_1=S_2=S_3=S_4=2$. 
Denoting the corresponding solution to \re{M-eq} as $\mathcal M_{TTTT}(x,\lambda)$, we deduce from 
\re{M-eq} that it should take the form of a homogenous polynomial in $Z_i^A$ of degree $4$, invariant under inversion and under the exchange of any pair of points. Examining \re{basis} we immediately find the solution
\begin{align}\label{TTTT-sol}
\mathcal M_{TTTT}(x,\lambda) \sim \vev{1\, 2\, 3\, 4}^4\,,
\end{align}
which agrees with  \re{TTTT-com}.
There exist other solutions, e.g. $\mathcal X_{1[234]}\mathcal X_{2[341]}\mathcal X_{3[412]}\mathcal X_{4[123]}/\prod_{i<j} x_{ij}^2$, but in contrast with \re{TTTT-sol} they have singularities for $x_{ij}^2\to 0$.  

The second example is the correlation function of two stress-energy tensors and two $1/2$ BPS operators, $S_1=S_2=2$ and
$S_3=S_4=0$. The corresponding function, $\mathcal M_{TTOO}(x,\lambda)$, is a homogenous polynomial in $Z_1$ and $Z_2$ 
of degree $2$. According to \re{M-eq}, it should have conformal weight $1/(x_3^2 x_4^2)^2$ under inversion and be invariant under the exchange 
of points $1\leftrightarrow 2$ and $3\leftrightarrow 4$. We use \re{basis} to obtain
\begin{align}\label{TTOO-sol}
\mathcal M_{TTOO}(x,\lambda) \sim \left(\mathcal X_{[12]3}\mathcal X_{[12]4}\right)^2= \big[ \vev{1|x_{13} x_{32}|2}\vev{1|x_{14} x_{42}|2}\big]^2\,.
\end{align}
As in the previous case, there are other solutions involving $\mathcal X_{1[234]}$ but we have to discard them since they are 
singular for $x_{ij}^2\to 0$. In distinction with \re{TTTT-sol}, the function $\mathcal M_{TTOO}$ should also depend on the variables $y$ at points $3$ and $4$. This  dependence  is unambiguously fixed by the $R-$symmetry and amounts to an additional factor of $(y_{34}^2)^2$ on the right-hand side of \re{TTOO-sol}. This factor has the meaning of the Clebsh-Gordon coefficient corresponding to the singlet representation in the tensor product decomposition ${\bf 20'\times 20'}$. %=\bf 1+ 15 + 20' + 84+105+175$.
 
The third example is the correlation function of two stress-energy tensors and two $R-$symmetry currents, $S_1=S_2=2$ and
$S_3=S_4=1$. Going along the same lines as in the two previous cases, we arrive at 
\begin{align}\label{TTJJ-sol}
\mathcal M_{TTJJ}(x,\lambda) \sim \vev{1\,2\,3\,4}^2  \mathcal X_{[12]3}\mathcal X_{[12]4} =\vev{1\,2\,3\,4}^2 \vev{1|x_{13} x_{32}|2}\vev{1|x_{14} x_{42}|2} \,.
\end{align}
The $y-$dependence of the $R-$current $J^{a_ia_i'}(x_i,y_i)$ leads to an  additional factor of $y_{34}^{a_3 a_4'} y_{34}^{a_4 a_3'}$ on 
the right-hand side of this relation. As in the previous case, it corresponds to the singlet representation in the tensor product
$\bf 15\times 15$. %=1+ 15_s + 15_a + 20'+45+\widebar{45} + 84$.

We notice that, up to an overall normalization factor, the obtained expressions \re{TTTT-sol}, \re{TTOO-sol} and \re{TTJJ-sol}
satisfy the following interesting relation
\begin{align}
(\mathcal M_{TTJJ})^2 \sim \mathcal M_{TTTT}\,\mathcal M_{TTOO}\,.
\end{align} 

\subsection{Comparison with known results}

Combined with \re{G-spins},  relations \re{TTOO-sol} and \re{TTJJ-sol} yield predictions for the correlation
functions involving two stress-energy tensors. The same correlation functions can be computed using the approach
described in section~\ref{sect:TTTT}. 

As an example, we can consider the correlation function  $\vev{T(1) T(2) O(3) O(4)}$ containing two $1/2$ BPS scalar 
operators. Following \re{G4-aux}, we identify the $\prod_{i=1,2} (\theta_i^2)^{\alpha_i\beta_i} (\bar\theta_i^2)_{\dot\alpha_i\dot\beta_i}$ component of \re{master}, and then apply the maximal number of derivatives recipe to 
 obtain a representation similar to \re{G4-der}. In close analogy with \re{subs}, the restoration of the conservation of the stress-energy tensor and its conformal properties can be achieved by symmetrizing the corresponding $\mathcal M-$tensor with respect to the Lorentz indices. Going along these lines we obtain
\begin{align}
\vev{T(1) T(2) O(3) O(4)} =   (\mathcal D_1\mathcal D_2)^2 \Big[(y_{34}^2)^2 \vev{1|x_{13} x_{32}|2}^2\vev{1|x_{14} x_{42}|2} ^2F(x)\Big].
\end{align}
In the same way, for the correlation function containing two $R-$symmetry currents we find
\begin{align}\notag
{}& \vev{T(1) T(2) J^{a_3a_3'}(3) J^{a_4a_4'}(4)} 
\\[2mm]
{}& \qquad =   (\mathcal D_1\mathcal D_2)^2\mathcal D_3\mathcal D_4 \Big[ y_{34}^{a_3 a_4'}y_{34}^{a_4 a_3'} \vev{1\,2\,3\,4}^2\vev{1|x_{13} x_{32}|2} \vev{1|x_{14} x_{42}|2} F(x)\Big],
\end{align}
where $J^{a_ia'_i}(i) = J^{a_ia'_i}_{\alpha\dot\alpha}(x_i) \lambda_i^\alpha \bar\lambda_i^{\dot\alpha}$.
Comparing these relations with \re{TTOO-sol} and \re{TTJJ-sol} we observe perfect agreement. 

We can apply the same technique to computing the correlation functions of other components of the stress-energy supermultiplet. The components $\mathcal T=\ldots + \theta^4   L(x) +\bar\theta^4  \bar L(x) +\dots $ are of particular interest since they define the chiral ($L$)  and antichiral ($\bar L$) on-shell Lagrangians of the theory,  with $\bar L(x) = L^\dagger (x)$ {and $i(L(x)-\bar L(x))$ being a topological term,}
\begin{align}\label{TTLL}\notag
{}& \vev{T(1) T(2) L(3) \bar L(4)} = (\mathcal D_1\mathcal D_2)^2 \Box_{x_4}^2 \left[\mathcal M_{TTL\bar L}(x,\lambda) F(x) \right],
\\[2mm]
{}& \vev{L(1) \bar L(2) L(3) \bar L(4)} = \Box_{x_2}^2\Box_{x_4}^2 \left[ (x_{24}^2)^4 F(x)\right],
\end{align}
with $\mathcal M_{TTL\bar L}(x,\lambda)= (\mathcal X_{[12]4})^4=\vev{1|x_{14}x_{42}|2}^4$. It easy to verify using \re{use}, \re{M-eq} and \re{inv-F} that the
expressions on the right-hand side of \re{TTLL} have the correct transformation properties under inversion. The second correlation function in \re{TTLL}  is equivalent to the result of \cite{Drummond:2006by} obtained by a different method. The latter has been used as a nontrivial consistency check of two ${\rm AdS_5 \times S^5}$ supergravity calculations, that  of the dilaton/axion amplitude in \cite{D'Hoker:1999pj} and  that for the bottom component of the same massive multiplet in \cite{Arutyunov:2000py}.
 
Another class of correlation functions containing three $1/2$ BPS operators and a conserved current has been studied
in \cite{Belitsky:2014zha}. We can use these results to obtain 
\begin{align}\notag
{}&  \vev{T(1) O(2) O(3) O(4)}  = \mathcal D_1 ^2 \left[y_{23}^2 y_{34}^2 y_{24}^2(\mathcal  X_{1[234]}(x,\lambda))^2 F(x)\right],
\\[2mm]
{}&  \vev{J^{aa'}(1) O(2) O(3) O(4)}   =\mathcal D_1\big[\mathcal Y_{1[234]}^{aa'}(y,x)\mathcal  X_{1[234]} (x,\lambda) F(x) \big],
\end{align}
where $\mathcal  X_{1[234]}$ was defined in \re{Xi} and the notation was introduced for 
\begin{align}
\mathcal Y_{1[234]} = 
x_{12}^2 x_{34}^2  y_{23}^2 y_{24}^2Y_{1[34]} + x_{23}^2 x_{14}^2 y_{24}^2y_{34}^2Y_{1[23]}
+ x_{13}^2 x_{24}^2  y_{23}^2 y_{34}^2  Y_{1[42]} 
\,, 
\end{align}
with $Y_{1[ij]}^{aa'} = (y_{1i} y_{ij} y_{j1})^{aa'}$. Here $\mathcal Y_{1[234]} $ and $Y_{1[ij]}^{aa'}$ carry the dependence on the $y-$variables and are completely antisymmetric under the exchange of the points indicated inside the  brackets. We recall
that the $R-$current and the $1/2$ BPS operator belong to the $SU(4)$ representations $\bf 15$ and $\bf 20'$, respectively. The three terms in the expression for $\mathcal Y_{1[234]}$ correspond to three overlapping $SU(4)$ representations
in the tensor products $\bf 15\times 20'$ and $\bf 20'\times 20'$. 

Thus, we demonstrated on various examples that the correlation functions of conserved currents in an $\mathcal N=4$ superconformal theory have the general form \re{G-spins} in which 
the information about the quantum numbers of the currents is encoded in the $\mathcal M-$function. In particular,
if the currents carry  $R-$charge, the $\mathcal M-$function can be decomposed into a sum of various irreducible 
components corresponding to overlapping $SU(4)$ representations in the different channels.

We would like to emphasize that the expressions for the correlation functions presented in this section are valid up to overall 
normalization factors. The latter can be determined, e.g., from the consistency with the operator product expansion.
 
\section{Application to energy-energy correlation}\label{sect:app}

In this section, we use our results for the four-point correlation functions to compute the energy-energy correlation
(EEC) in an $\mathcal N=4$ superconformal
theory. This infrared safe observable describes the flow of energy in the final state created from the vacuum by some source
\cite{Basham:1978bw}. More precisely, ${\rm EEC}(n_1,n_2)$ 
measures the correlation between the energy fluxes in two different directions defined by light-like four-vectors $n_1$ and $n_2$ (with $n_1^2=n_2^2=0$). As such, it is expected to be a regular positive-definite function of the angles defining the relative orientation of $n_1$ and $n_2$.

\subsection{Generalized optical theorem}

The energy-energy correlation admits the following representation in terms of a correlation function
\cite{Ore:1979ry,Sveshnikov:1995vi,Korchemsky:1997sy,Korchemsky:1999kt,Belitsky:2001ij}
\begin{align}\notag\label{EEC-ini}
{}& {\rm EEC}(n_1,n_2) = \sigma^{-1} \int d^4 x \e^{i(x q)} G(x;n_1,n_2)\,,
\\
{}& G(x_{3}-x_4;n_1,n_2) = \vev{\mathcal E(n_1) \mathcal E(n_2) J(x_3) \widebar J(x_4)}_W\,,
\end{align}  
where the Fourier integral fixes the total momentum of the final state to be $q$. The normalization factor $\sigma$ 
is fixed by the requirement that, in the rest frame of the source, for  $q=(E,\vec 0)$ and $n_i=(1,\vec n_i)$, the energy-energy correlation averaged over the spatial orientation of the unit vectors $\vec n_i$ has to satisfy the condition
$\int d\Omega_{\vec n_1} d\Omega_{\vec n_2} {\rm EEC}(n_1,n_2) =1$.
 
The correlation function $G(x_{34};n_1,n_2)$ involves the so-called energy flow operators, $\mathcal E(n_1)$ and $\mathcal E(n_2)$, which are expressed in terms of integrated
stress-energy tensors (see \cite{Sveshnikov:1995vi,Korchemsky:1997sy} for the explicit expression). The operator  $J(x)$ defines the source. If the source is created by a conserved current of spin $S$, the operator $J(x)$ depends on 
the polarization tensor $e$
\begin{align}\label{source}
J(x) =  e^{\alpha_1 \dots \alpha_S}_{\dot\alpha_1\dots  \dot\alpha_S} \,J_{\alpha_1 \dots \alpha_S}^{\dot\alpha_1\dots  \dot\alpha_S}(x)\,.
\end{align}
The operator $\widebar J(x)= J^\dagger(x)$ is given by a similar expression with $e$ replaced by the conjugated polarization  tensor $\bar e$.
The subscript $\scriptstyle W$ in the second line of \re{EEC-ini} indicates that this is the Wightman (not time-ordered) four-point function. It is related to 
its Euclidean counterpart via analytic continuation.

To compute \re{EEC-ini} it proves convenient to perform a conformal transformation and go to new coordinates $x^\mu \to z^\mu$
\cite{Cornalba:2007fs,Hofman:2008ar}
\begin{align}\label{z-var}
z^+ = - {1\over x^+}\,,\qqquad z^-=x^- - {x\bar x \over x^+}\,,\qqquad \vec z={\vec x\over  x^+}\,,\qqquad dz^2 = {dx^2\over (x^+)^2} \,,
\end{align}
where the notation was introduced for the light-like coordinates $x^\pm$ and $\vec x= (x,\bar x)$,
\begin{align}\label{LC}
x_{\alpha\dot\alpha} = x_\mu (\sigma^\mu)_{\alpha\dot\alpha}
= \bigg[\begin{array}{ll} \ x^+ & \bar x \\ \ x & x^- \end{array}\bigg]\,,\qqqquad dx^2 = dx^+ dx^- - dx \,d \bar x\,.
\end{align} 
Then, if the source \re{source} is defined by an operator of spin $S$ and dimension $\Delta_S$, the function $G(x_{34};n,n')$ in \re{EEC-ini} is given by the four-point correlation function  integrated  over the light-cone coordinates of the two stress-energy tensors
\begin{align}\notag\label{G-int}
G_S(x_{34};n_1,n_2) {}& = {(z_3^+ z_4^+)^{\Delta_S -S}\over (n_1^+ n_2^+)^3}e^{\alpha_1 \dots \alpha_S}_{\dot\alpha_1\dots  \dot\alpha_S}\,
\bar e^{\beta_1 \dots \beta_S}_{\dot\beta_1\dots\dot\beta_S}\Lambda_{\alpha_1 \dot\gamma_1}^{\dot\alpha_1\gamma_1}(3)
\Lambda_{\beta_1 \dot\delta_1}^{\dot\beta_1\delta_1}(4) 
\dots
\Lambda_{\alpha_S\dot\gamma_S}^{\dot\alpha_S\gamma_S}(3) 
\Lambda_{\beta_S \dot\delta_S}^{\dot\beta_S\delta_S}(4) 
\\[2mm]
{}& \times \int_{-\infty}^\infty dz_1^- dz_2^- \vev{T(0^+,z_1^-,\vec z_1) T(0^+,z_2^-,\vec z_2) J_{\gamma_1 \dots \gamma_S}^{\dot\gamma_1\dots  \dot\gamma_S}(z_3)
 J_{\delta_1 \dots \delta_S}^{\dot\delta_1\dots\dot\delta_S}(z_4)}_W\,,
\end{align}
where the product of  the $z-$dependent factor and  $\Lambda_{\alpha \dot\gamma}^{\dot\alpha\gamma}(i)=\partial z^{\gamma}_{i,\dot\gamma}/ \partial x^{\alpha}_{i,\dot\alpha}$ in the first line arises from  the conformal transformation \re{z-var} of the two currents. Here
$T(z_i^+=0,z_i^-,\vec z_i)$ (with $i=1,2$) are the stress-energy tensors \re{T-pro} whose Lorentz indices are projected onto the same auxiliary spinors
$\lambda_0^\alpha=(1,0)$ and $\bar\lambda_0^{\dot\alpha}=\left({1\atop 0}\right)$ satisfying $\lambda_0^\alpha z_{\alpha\dot\alpha} \bar\lambda_0^{\dot\alpha} = z^+$. 
The nontrivial dependence of \re{G-int} on the light-like vectors $n_1$ and $n_2$ comes through the variable  $\vec z_i = \vec n_i/n_i^+$, depending on the light-cone coordinates of these vectors.

According to \re{EEC-ini} and \re{G-int}, the definition of the energy-energy correlation depends on the choice of the source $J$.
Later in this section, we consider four choices for $J$: a $1/2$ BPS operator, an $R-$symmetry current,
a stress-energy tensor and a Lagrangian of the theory.\footnote{In all cases except the last one we have $\Delta_S-S=2$.  The Lagrangian appears as $O(\theta^4)$ component of the stress-energy supercurrent $\mathcal T$ and it has dimension $4$ and spin zero.} We will show that, in virtue of $\mathcal N=4$ superconformal symmetry, the energy-energy correlations are given  in all four cases 
by the same expression and, therefore, do not depend on the choice of the source. 

\subsection{Integrated correlation functions}

Let us first examine the Euclidean version of the correlation function entering \re{G-int}. 
In  close analogy with \re{T4-sum}, this correlation function can be split into a sum of rational and anomalous pieces. As 
was shown in \cite{Belitsky:2013xxa,Belitsky:2013bja}, the contribution of the former to the energy-energy correlation \re{EEC-ini} vanishes for generic $n_1$ and $n_2$ 
and, therefore, can be discarded.\footnote{More precisely, it is proportional to a delta function with support  at 
$\chi=q^2 (n_1n_2)/(2 (qn_1)(qn_2))=1$.}
 
The anomalous contribution to the correlation function in \re{G-int} can be found from \re{G-spins} for $S_1=S_2$ and $S_3=S_4=S$. More precisely, in order to match \re{G-int}, we have
to identify the auxiliary spinors, $\lambda_1=\lambda_2=\lambda_0$  and $\bar\lambda_1=\bar\lambda_2=\bar\lambda_0$. Taking into
account that $\lambda_0$ and $\bar\lambda_0$ satisfy $\lambda_0^\alpha z_{\alpha\dot\alpha} \bar\lambda_0^{\dot\alpha} = z^+$ for an arbitrary four-vector $z$, we can simplify the expression for the differential operators $\mathcal D_1$ and $\mathcal D_2$
defined in \re{Di} (see \re{aux-eq}) \footnote{Notice that we have to apply to derivative $\partial_{\lambda_i}$ before putting $\lambda_i=\lambda_0$.}
\begin{align}
\mathcal D_i = \bar\lambda_{0,\dot\alpha} (\partial_{z_i})^{\dot\alpha\alpha}(\partial_{\lambda_i})_{\alpha} = -  
\lambda_0^\alpha (\partial_{\lambda_i})_{\alpha}\partial_{\bar z_i} - \partial_{z_i^-}\,.
\end{align}
In addition, we can safely neglect terms containing $\partial_{z_i^-}\equiv \partial/\partial z_i^-$ since they produce a contribution 
to the correlation function \re{G-spins} that integrates to zero after substitution in \re{G-int}. Since the expression inside the 
brackets in \re{G-spins} is a homogenous function of $\lambda_i$, we can replace $\mathcal D_i\sim \partial_{\bar z_i}$ for $i=1,2$. In this way, we obtain from  \re{G-spins}
\begin{align}\label{M-dots}
\vev{T(1) T(2) J_S(3) J_S(4) } = (\partial_{\bar z_1}\partial_{\bar z_2})^2  (\mathcal D_3 \mathcal D_4)^{S}\Big[ \mathcal M_{TTJ_SJ_S}(z,\lambda)  F(z)\Big] +\dots\,,
\end{align}
where the dots denote terms containing total derivatives with respect to $z_1^-$ and $z_2^-$, and  
we used the shorthand notation for $T(i)=T(0^+,z_i^-,\vec z_i)$ and
\begin{align}
J_S(k)=\lambda_{k}^{\gamma_1}\bar\lambda_{k,\dot\gamma_1}\dots \lambda_{k}^{\gamma_S}\bar\lambda_{k,\dot\gamma_S} J_{\gamma_1 \dots \gamma_S}^{\dot\gamma_1\dots  \dot\gamma_S}(z_k)\,.
\end{align}
For a source defined by the Lagrangian of the theory, we find from \re{TTLL} that the correlation function \re{M-dots} takes a slightly different form
\begin{align} \label{M-dots1}
\vev{T(1) T(2) L(3) \bar L(4) } = (\partial_{\bar z_1}\partial_{\bar z_2})^2 \,  \Box_{z_4} ^{2}\Big[ \mathcal M_{TTL\bar L}(z,\lambda)  F(z)\Big] +\dots\ .
\end{align}
As before, to obtain  the correlation function in \re{G-int}, it suffices to differentiate both sides of \re{M-dots}  with respect to the auxiliary spinors $\lambda_k$ and $\bar\lambda_k$ with $k=3,4$.

At the next step, we use the expressions for $\mathcal M_{TTJ_SJ_S}$ obtained in the previous section to evaluate them in terms of the
$z-$coordinates. For instance, for the $1/2$ BPS  operator $O_{\bf 20'}$ we apply \re{TTOO-sol} to get (with $z_{ij}=z_i-z_j$)
\begin{align}\label{Mz1}
\mathcal M_{TTOO} ={}& (y_{34}^2)^2\big[ \vev{0|z_{13} z_{32}|0}\vev{0|z_{14} z_{42}|0}\big]^2 =(y_{34}^2)^2 \bar z_{12}^4 (z_3^+ z_4^+)^2\,.
\end{align} 
Here in the first relation we rewrote \re{TTOO-sol} in $z-$coordinates, added the $y-$dependent factor and identified the 
auxiliary spinors $\ket{1}=\ket{2}=\ket{0}$. In the second relation, we replaced $(z_i)_{\alpha\dot\alpha}$ by its expressions \re{z-var} and \re{LC}
in terms of light-cone variables and took into account that $z_1^+=z_2^+=0$.

For the Lagrangian, the stress-energy tensor and the $R-$current, we apply relations \re{TTLL},  
 \re{TTTT-sol} and \re{TTJJ-sol}, respectively, and go through similar calculations to find
\begin{align}\notag\label{Mz2}
{}& \mathcal M_{TTL\bar L} = \bar z_{12}^4 (z_4^+)^4\,,
\\[2mm]\notag
{}& \mathcal M_{TTTT} =  \bar z_{12}^4 \vev{3|\omega| 4}^4\,,
\\[1.2mm]
{}& \mathcal M_{TTJJ} =y_{34}^{a_3a_4'} y_{34}^{a_4a_3'}  \bar z_{12}^4 z_3^+ z_4^+ \vev{3|\omega| 4}^2\,,
\end{align}
where $ \vev{3|\omega| 4} = \lambda_3^\alpha\, \lambda_4^{\beta} \,\omega_{\alpha \beta}$ and the matrix $\omega_{\alpha \beta}$ is given by 
\begin{align}\label{omega}
\omega=  \left[\begin{array}{cc}0 & z^+_3 \\ -z^+_4 & z_{34} \end{array}\right]\,.
\end{align}

Examining \re{Mz1} and \re{Mz2} we observe an interesting property: the obtained expressions for the $\mathcal M-$functions
are independent of the coordinates $z_1^-$ and $z_2^-$ and, at the same time, they have the same dependence on $\bar z_{12}$.
The former property implies that, upon substitution of \re{M-dots} and \re{M-dots1} into \re{G-int}, the integration over $z_1^-$ and $z_2^-$  can be reduced to evaluating the following
integral
\begin{align}\label{cal-G}
\int_{-\infty}^\infty dz_1^- dz_2^-\, F(z) = {1\over   (\vec z_{12}^{\,2} z_{34}^2 )^3 z_3^+ z_4^+} \mathcal G(\gamma)\,,
\end{align}
where the function $F(z)$ was defined in \re{F-Phi}.
Here the $z-$dependent factor on the right-hand side carries the scaling dimension of the integral, so that $\mathcal G(\gamma)$ is dimensionless. Moreover, the detailed analysis shows \cite{Belitsky:2014zha,Belitsky:2013xxa,Belitsky:2013bja} that the argument of  $\mathcal G(\gamma)$ is a rational function of $\gamma=\gamma(z_i)$ whose form is much simpler when expressed in terms of $x-$coordinates
\begin{align}
\gamma =  2{(x_{34} n_1) (x_{34} n_2)\over x_{34}^2 (n_1 n_2)}\,.
\end{align}
The explicit expressions for $\mathcal G(\gamma)$ in $\mathcal N=4$ SYM at weak and at strong coupling can be found in \cite{Belitsky:2013xxa}.
We do not need them for our purposes.

\subsection{EEC for a scalar source}

To avoid the technical difficulties related to the Lorentz structure of the source, 
let us first consider \re{G-int} with a scalar source $J$ given by the $1/2$ BPS operator $O_{\bf 20'}$ or the Lagrangian $L$. We apply 
\re{M-dots} -- \re{Mz2} and \re{cal-G} to get for   $\Delta_{O_{\bf 20'}}=2$, $\Delta_L=4$ and $S=0$
\begin{align}\notag\label{GO}
G_O(x_{34};n_1,n_2) {}&
={(y_{34}^2)^2}{ (z_3^+z_4^+)^3\over (n_1^+ n_2^+\vec z_{12}^{\,2} z_{34}^2 )^3}
   (\bar z_{12})^3 (\partial_{\bar z_1}\partial_{\bar z_2})^2   [ \bar z_{12}\,\mathcal G(\gamma)],
\\
G_L(x_{34};n_1,n_2) {}&
= (z_4^+)^4 \Box_{z_4}^2{(z_3^+z_4^+)^3 \over (n_1^+ n_2^+ \vec z_{12}^{\,2} z_{34}^2)^3}(\bar z_{12})^3(\partial_{\bar z_1}\partial_{\bar z_2})^2   \left[{ \bar z_{12}}\,\mathcal G(\gamma)   \right],
\end{align}
where the subscript on the left-hand side indicates the choice of the source. 
We can now use \re{z-var} to write the expressions on the right-hand side in terms of $x_i$. This can be done with the help of the 
identity 
\begin{align}\label{id}
 \left[\gamma^2(1-\gamma)^2 
\mathcal G ''(\gamma)\right]'' =
(\bar z_{12})^3(\partial_{\bar z_1}\partial_{\bar z_2})^2   \left[  \bar z_{12}\,\mathcal G(\gamma)   \right] = {1\over 16} (x_{34}^2)^3 \,\Box_{x_3}^2 {\mathcal G(\gamma)\over x_{34}^2}\,,
\end{align} 
which can be verified by a straightforward calculation. Taking into account that $2(n_1n_2)=n_1^+ n_2^+ \vec z_{12}^{\,2}$ and $x_{34}^2=z_{34}^2/(z_3^+z_4^+)$ we arrive at Lorentz covariant expressions for the integrated correlation functions 
\begin{align}\notag\label{O-L}
G_O(x_{34};n_1,n_2) {}& = {(y_{34}^2)^2\over 128(n_1 n_2)^3}\Box_{x_3}^2 {\mathcal G(\gamma)\over x_{34}^2}\,,
\\
G_L(x_{34};n_1,n_2) {}& =  {1 \over 128(n_1 n_2)^3}\Box_{x_3}^2\Box_{x_3}^2 {\mathcal G(\gamma)\over x_{34}^2}\,.
\end{align}
We notice that the two expressions are related to each, $G_L\sim \Box_{x_3}^2 G_O$.

To compute the energy-energy correlation \re{EEC-ini}, we have to analytically continue relations \re{O-L} to
get the Wightman correlation functions, and then Fourier transform  them with respect to 
$x_{34}=x_3-x_4$. It is easy to see that the Fourier transforms of $G_O$ and $G_L$ are proportional to the same dimensionless
function, e.g.
\begin{align}
 \int d^4 x \e^{ix q}G_L(x;n_1,n_2) = {(q^2)^3 \over 128(n_1 n_2)^3}\mathcal F(\chi)\,,
\end{align}
where the notation was introduced for $\chi={q^2 (n_1n_2)/( 2(qn_1)(qn_2))}$ and
\begin{align}\label{cal-F}
\mathcal F(\chi) = q^2 \int d^4 x \, {\rm e}^{i q x} {\mathcal G_W(\gamma)\over x^2 -i0 x^0} \,.
%\,,\qqqquad z={q^2 (n_1n_2)\over 2(qn_1)(qn_2)}\,.
\end{align} 
Here  $\mathcal G_W(\gamma)$ corresponds to the particular analytic continuation of its Euclidean counterpart $\mathcal G(\gamma)$. Together with the  `$-i0 x^0$' prescription in the denominator this ensures that the Fourier integral \re{cal-F} is different
from zero only for $q^2>0$ and $q_0>0$, as it should be for a physical quantity measuring the flow of energy in a final
state with total momentum $q$. As a consequence, $\mathcal F(\chi)$ is different from zero for $0<\chi < 1$. In the
rest frame of the source, for $q=(E,\vec 0)$ and $n_i=(1,\vec n_i)$ (with $\vec n_i^2=1$), the scaling variable $\chi=(n_1n_2)/2=(1-\cos\theta)/2$ is related 
to the angle $\theta$ between the unit vectors $\vec n_1$ and $\vec n_2$.

Substituting \re{O-L} into \re{EEC-ini} we find that the energy-energy correlations in both cases are proportional to $\mathcal F(\chi)$. The proportionality
factor can be determined by imposing the normalisation condition 
\begin{align}\label{EEC-norm}
\int_0^1 d\chi\, {\rm EEC}(\chi) = \frac12\,,
\end{align}
which follows from the requirement for the total energy in the final state to be equal to the momentum transferred $q$. In this way,
we arrive at
\begin{align}\label{uni}
{\rm EEC}_{O} = {\rm EEC}_{L} = {\mathcal F(\chi)  \over \chi^3}\,, %\qqqquad \chi={q^2 (n_1n_2) \over 2(qn_1)(qn_2)}\,.
\end{align}
with $\chi=(n_1n_2)/2=(1-\cos\theta)/2$ in the rest frame of the source.
 
\subsection{EEC for a tensor source}

For a scalar source,   Lorentz invariance implies that ${\rm EEC}(n_1,n_2)$ can only depend on the relative angle $\chi$ 
between $n_1$ and $n_2$. For a source defined by the $R-$current and the stress-energy tensor, this is not necessarily the case
due to the dependence of ${\rm EEC}(n_1,n_2)$ on the polarization vectors \re{source}. Nevertheless, as we show in this
subsection, the energy-energy correlations in an $\mathcal N=4$ superconformal theory do not depend on the choice of the source
and are given by the universal scaling function \re{uni}.

For a source given by a current of spin $S$, an additional complication arises in \re{G-int}
due to the necessity to deal with Lorentz indices. Let us first examine the expression in the second line of \re{G-int} with 
the four-point correlation function replaced by \re{M-dots}. According to \re{Mz2},  the $\mathcal M-$functions corresponding
to the $R-$current and stress-energy tensor do not depend on $z_{1,2}^-$ and admit the following representation
$\mathcal M_{TTJ_SJ_S}\sim  \bar z_{12}^4 \vev{3|\omega| 4}^{2S}(z_3^+ z_4^+)^{2-S}$ for $S=1$ and $S=2$, respectively.
%, and $\Delta_S-S=2$.
Then, we find from \re{M-dots}  
\begin{align}\notag\label{int-gen}
  \int_{-\infty}^\infty dz_1^- dz_2^-\, {}&\vev{T(1) T(2) J_S(3) J_S(4) }
\\\notag
 {}& \sim   (\mathcal D_3 \mathcal D_4)^{S} (z_3^+ z_4^+)^{2-S} \vev{3|\omega| 4}^{2S}(\partial_{\bar z_1}\partial_{\bar z_2})^2  \bar z_{12}^4 \int_{-\infty}^\infty dz_1^- dz_2^-\, F(z) 
\\
{}& = (\mathcal D_3 \mathcal D_4)^{S} {\vev{3|\omega| 4}^{2S}\over  (z_3^+ z_4^+)^{S+2}}{  (z_3^+ z_4^+)^{3}\over (\vec z_{12}^{\,2} z_{34}^2 )^3  }
 (\bar z_{12})^3(\partial_{\bar z_1}\partial_{\bar z_2})^2 \big[ \bar z_{12} \,
   \mathcal G(\gamma)\big]\,,
\end{align}
where in the second relation we applied \re{cal-G}. 

We observe the striking similarity of \re{int-gen} with the first relation in \re{GO}. An important
difference is however that the last relation involves the derivatives $\mathcal D_{3,4}$ and the matrix element $\vev{3|\omega|4}$ defined in \re{Di} and \re{omega}, respectively. They induce the dependence of \re{int-gen} on the auxiliary spinors $\lambda_i$ and $\bar\lambda_i$ with $i=3,4$. To obtain the correlation function in the second line of \re{G-int}, we have to differentiate \re{int-gen}
with respect to $\lambda_i$ and $\bar\lambda_i$. In this way, we arrive at the following differential operator
\begin{align}\notag \label{P-op}
 {}&P_S(\partial_{x_3},\partial_{x_4})  =
(z_3^+ z_4^+)^{2}e^{\alpha_1 \dots \alpha_S}_{\dot\alpha_1\dots  \dot\alpha_S}\,
\bar e^{\beta_1 \dots \beta_S}_{\dot\beta_1\dots\dot\beta_S}
\\ %\notag
{}&\qquad  \times \prod_{i=1}^S \Lambda_{\alpha_i \dot\gamma_i}^{\dot\alpha_i\gamma_i}(3)
\Lambda_{\beta_i \dot\delta_i}^{\dot\beta_i\delta_i}(4) 
 (\partial_{\lambda_3})_{\gamma_i}  
  (\partial_{\lambda_4})_{\delta_i}
  (\partial_{\bar\lambda_3})^{\dot\gamma_i}
   (\partial_{\bar\lambda_4})^{\dot\delta_i}\,(\mathcal D_3 \mathcal D_4)^{S} {\vev{3|\omega| 4}^{2S}\over  (z_3^+ z_4^+)^{S+2}}\,,
\end{align}
where $\Lambda_{\alpha \dot\gamma}^{\dot\alpha\gamma}(i)=\partial z^{\gamma}_{i,\dot\gamma}/ \partial x^{\alpha}_{i,\dot\alpha}$ and $\mathcal D_i=\bar\lambda_{i,\dot\alpha} (\partial_{z_i})^{\dot\alpha\alpha}(\partial_{\lambda_i})_{\alpha}$. It is easy to see
that it does not depend on the auxiliary spinors and is given by a linear combination of powers of the differential operators 
$\partial_{z_3}$ and $\partial_{z_4}$. Using \re{z-var} they can be converted   into the differential operators $\partial_{x_3}$ and $\partial_{x_4}$. For $S=1,2$ the explicit expressions for $P_S(\partial_{x_3},\partial_{x_4})$ are given below in \re{P-pol}. Notice that
\re{P-op} only depends on points $3$ and $4$. % and is independent of $n_1$ and $n_2$.

Combining together \re{int-gen} and \re{P-op} we find from \re{G-int} with the help of the identity \re{id}  
\begin{align}\label{P-diff}
G_S(x_{34};n_1,n_2) = {1\over 128(n_1 n_2)^3} P_S(\partial_{x_3},\partial_{x_4}) \,\Box_{x_3}^2 {\mathcal G(\gamma)\over x_{34}^2}\,.
\end{align}
We observe that the only difference with \re{O-L} is the appearance of the differential operator $P_S(\partial_{x_3},\partial_{x_4})$
that carries the dependence on the polarization tensor of the current. Upon the Fourier transform \re{EEC-ini}, this operator is
replaced by $P_S(q,-q)$
\begin{align}\label{G-uni}
 \int d^4 x \e^{ix q}G_S(x;n_1,n_2) = {(q^2)^{3} \over 128(n_1 n_2)^3}\,P_S(q,-q) \,\mathcal F(\chi)\,,
\end{align}
where the function $\mathcal F(\chi)$ is given by \re{cal-F}.
Going through a lengthy calculation we find from  \re{P-op}
\begin{align}\notag\label{P-pol}
P_{S=1}(q,-q) {}&= q^2e^{\alpha}_{\dot\alpha}\bar e^{\dot\alpha}_{\alpha} +\frac12  (e^{\alpha}_{\dot\alpha} q_{\alpha}^{\dot\alpha}) (\bar e^{\beta}_{\dot\beta}q_{\beta}^{\dot\beta}) 
 = 2 q^2 e^\mu \bar e_\mu + 2 (e^\mu q_\mu) (\bar e_\nu q^\nu)\,,
\\[2mm] \notag
P_{S=2}(q,-q){}& =6 \big(e^{\alpha_1\beta_1}_{\dot\alpha_1\dot\beta_1} q_{\alpha_1}^{\dot\alpha_1} q_{\beta_1}^{\dot\beta_1} \big)\big(\bar e^{\alpha_2\beta_2}_{\dot\alpha_2\dot\beta_2}  
q_{\alpha_2}^{\dot\alpha_2} q_{\beta_2}^{\dot\beta_2}  \big)
 + 36 q^2  \big( e^{\alpha_1\beta_1}_{\dot\alpha_1\dot\beta_1} q_{\beta_2}^{\dot\beta_1} \big)  \big(\bar e^{\alpha_2\beta_2}_{\beta_2\dot\beta_2}   
 q_{\alpha_1}^{\dot\alpha_2}\big)\epsilon^{\dot\alpha_1\dot\beta_2}\epsilon_{\beta_1\alpha_2}
 \\[2mm]
{}&= 24\big[ (e^{\mu\nu}q_\mu q_\nu)(\bar e_{\mu\nu}q^\mu q^\nu) + 12 q^2  e^{\mu\rho}\bar e_{\rho\nu} q_\mu q^\nu-  6 (q^2)^2 e^{\mu\nu}\bar e_{\mu\nu}\big],
\end{align}
where $e^\mu= \frac12 e^\alpha_{\dot\alpha} (\sigma^\mu)_\alpha^{\dot\alpha}$ and $e^{\mu\nu}= \frac14 e^{\alpha\beta}_{\dot\alpha\dot\beta} (\sigma^\mu)_\alpha^{\dot\alpha}(\sigma^\nu)_\beta^{\dot\beta}$ are the polarization tensors for currents of spin $S=1$ and $S=2$, respectively.

The very fact that the expression on the right-hand side of \re{G-uni} factorizes into a product of the polynomial $P_S(q,-q)$ and the universal scaling function $\mathcal F(\chi)$ immediately implies that the energy-energy correlations \re{EEC-ini} normalized according to \re{EEC-norm} cease to depend on the spin of the current and the polarization tensor. We therefore conclude that in $\mathcal N=4$ superconformal theory
the energy-energy correlations do not depend on the choice of the source 
\begin{align}\label{uni1}
{\rm EEC}_{J} = {\rm EEC}_{T}={\rm EEC}_{O} = {\rm EEC}_{L} = {\mathcal F(\chi)  \over \chi^3}\,.
\end{align}
In the $\mathcal N=4$ SYM theory, the function $\mathcal F(\chi)$ is known at next-to-leading order both at weak \cite{Belitsky:2013ofa} and at strong coupling \cite{Hofman:2008ar,Belitsky:2013xxa,Goncalves:2014ffa}. 

\section{Conclusions}\label{sect:conc}

In this paper, we have studied the four-point correlation functions of the conserved currents in $\mathcal N=4$ superconformal theory. Contracting the Lorentz indices of the currents with  
auxiliary spinors we found that the correlation functions have a remarkably simple form -- they are given by total spatial derivatives acting on some scalar functions, with the 
number of derivatives related to the spin of the currents. The scalar functions factor out into the product of a universal dynamical function and kinematical $\mathcal M-$functions
depending on the chiral auxiliary spinors and the space-time coordinates of the operators. We demonstrated that the requirement for the correlation functions to respect the conservation of the 
currents and to have correct conformal properties lead  to powerful constraints on the $\mathcal M-$functions. This allowed us to determine them for various
correlation functions without doing any calculations. 

The obtained results for the $\mathcal M-$functions revealed a surprising similarity with the known expressions for the scattering amplitudes in $\mathcal N=4$ SYM theory.
Promoting the auxiliary spinors to new, {twistor-like} coordinates of the operators and assigning them a definite conformal weight, we found that the $\cal M-$functions are built
from very special blocks which have been previously identified as the simplest, four-point dual conformal invariants for the scattering amplitudes. We believe that 
this is not accidental and hints at the existence of an additional symmetry of the correlation functions. {Our results are in agreement 
with the recent findings of \cite{Chicherin:2014uca} for a new class of $\mathcal N=4$ superconformal invariants. They admit a compact
representation if expressed in terms of twistor variables analogous to \re{Z} and their linear combinations describe the multiple-point correlation functions of $\mathcal N=4$ supercurrents in the chiral sector. It would be interesting to generalize such a representation 
to include the dependence on the anti-chiral variables $\bq$. }
  
It is natural to ask whether the higher-point correlation functions of the stress-energy tensor admit the same `derivative' representation as the four-point functions. It is interesting to note that a similar
representation exists for the two-point function $\vev{T(1) T(2)}\sim (\mathcal D_1\mathcal D_2)^2\left[ \vev{12}^4/(x_{12}^2)^2\right]$, but not for the three-point function since otherwise it would imply the vanishing of the two-point  function, in virtue of the Ward identities \cite{Erdmenger:1996yc}. We 
recall that the distinguishing feature of the derivative representation is that the resulting expression for the correlation function satisfies the homogenous Ward identities \re{WI}. In general,
the contact terms in the Ward identities are proportional to correlation functions with fewer points. Then, the absence of contact terms for the anomalous contribution
to the four-point function is an immediate consequence of the protectedness of the two- and three-point functions in an $\mathcal N=4$ superconformal theory. Since the four-point function is not
protected, we do not expect the higher-point correlation functions to obey homogenous Ward identities, thus making the derivative representation problematic.  

One of the byproducts of our analysis is the prediction for the four-point correlation function of stress-energy tensors in planar $\mathcal N=4$ SYM theory at strong coupling. It
is given by \re{T4-sum},  \re{TTTT-com} and \re{F-Phi} with the function $\Phi(u,v)$ replaced by its expression found in \cite{Arutyunov:1999fb,Arutyunov:2000py,Arutyunov:2000ku}.
Via the AdS/CFT correspondence this correlation function is dual to the four-graviton scattering amplitude in ${\rm AdS_5}$. Due to the complexity of the corresponding Witten diagrams, such an amplitude has not been computed so far. The fact that the correlation function has the special form described above should simplify the problem 
and make the calculation feasible. 

\section*{Acknowledgments}

We are grateful to Omer Gurdogan for collaboration on an early stage of this project. We are indebted to  Andrei
Belitsky, Stefan Hohenegger and Sasha Zhiboedov  for previous collaboration on related topics.
E.S. thanks  Dima Chicherin, Hugh Osborn, Yassen Stanev and Ivan Todorov  for interesting discussions. 
We acknowledge partial support by
the French National Agency for Research under contract BLANCSIMI-4-2011.
G.K. would like to thank  the Galileo Galilei Institute for Theoretical Physics
and
FAPESP grant 2011/11973-4 for funding his visit to ICTP-SAIFR where part of this work was done.

\appendix 

\section{Conventions}\label{App-conv}

A four-dimensional vector $x_\mu$ can be represented by a $2\times 2$ matrix
\begin{align}
x_{\alpha\dot\alpha} = x_\mu (\sigma^\mu)_{\alpha\dot\alpha} = \bigg[\begin{array}{ll}\ x^+ & \bar x \\ \ x & x^-\end{array}\bigg]\,,\qqqquad
x_\mu^2=\det \| x_{\dot\alpha\alpha}\| = x^+ x^- - x \,\bar x\,.
\end{align}
We use the following conventions for rising/lowering indices
\begin{align}  
x^\alpha_{\dot\alpha} = \epsilon^{\alpha\beta} x_{\beta\dot\alpha} \,,\qquad 
x_\alpha^{\dot\alpha} = x_{\alpha\dot\beta}  \epsilon^{\dot\beta\dot\alpha} \,,\qquad 
 x^{\dot\alpha\alpha} =\epsilon^{\alpha\beta} x_{\beta\dot\beta}\epsilon^{\dot\beta\dot\alpha} =  \bigg[\begin{array}{cc}-x^- & \bar x \\ x & -x^+\end{array}\bigg],
\end{align} 
where the completely antisymmetric tensors are normalized as $\epsilon^{12} = \epsilon_{12}=\epsilon^{\dot 1\dot 2} = \epsilon_{\dot 1\dot 2}=1$ and satisfy the relations  
\begin{align}
\epsilon^{\alpha\beta}\epsilon_{\gamma\beta} =\delta^\alpha_\gamma\,,\qqqquad 
\epsilon^{\dot \alpha\dot \beta}\epsilon_{\dot \gamma\dot \beta} = \delta^{\dot \alpha}_{\dot \gamma}\,.
\end{align}
Defining the derivatives by $(\partial_x)^{\dot\alpha\alpha} x_{\beta\dot\beta} = \delta^\alpha_\beta \delta^{\dot\alpha}_{\dot\beta}$, we find for  $\bar\lambda_{0,\dot\alpha} =(0,-1)$
\begin{align} \label{aux-eq}
(\partial_x)^{\dot\alpha\alpha} =  \bigg[\begin{array}{ll}\ \partial_{x^+} & \partial_x \\ \ \partial_{\bar x} & \partial_{x^-}\end{array}\bigg]\,,
\qqqquad
\bar\lambda_{0,\dot\alpha}(\partial_x)^{\dot\alpha\alpha} = -\bigg(\begin{array}{l} \partial_{\bar x} \\ \partial_{x^-} \end{array}\bigg)\,.
\end{align}
It is straightforward to verify that  under inversion $I[x_{\alpha\dot\beta}]=(x^{-1})_{\beta\dot\alpha}=x_{\beta\dot\alpha}/x^2$ the derivatives transform  as
\begin{align}\notag\label{invs}
{}&   I\big[ \partial^{\dot\alpha_1 \beta_1} \big]    
=  (x^2)^2   x^{\alpha_1}_{\dot\gamma_1}  \partial^{\dot\gamma_1\gamma_1}   x_{\gamma_1}^{\dot\beta_1}\,{1\over (x^2)^2} 
=  x^{\alpha_1}_{\dot\gamma_1}  x_{\gamma_1}^{\dot\beta_1} \partial^{\dot\gamma_1\gamma_1 }   \,,
\\\notag
{}& I\big[ \partial^{\dot\alpha_1\beta_1} \partial^{\dot\alpha_2\beta_2}\big] =  (x^2)^2 
x^{\alpha_1}_{\dot\gamma_1} \, x^{\alpha_2}_{\dot\gamma_2}  \, (\partial^{\dot\gamma_1\gamma_1}\partial^{\dot\gamma_2\gamma_2} )
x_{\gamma_1}^{\dot\beta_1}x_{\gamma_2}^{\dot\beta_2'}\, {1\over (x^2)^2}\,,
\\ 
{}&  I\big[ \partial^{\dot\alpha_1\beta_1} \partial^{\dot\alpha_2\beta_2} \partial^{\dot\alpha_3\beta_3}\big] = (x^2)^2
x^{\alpha_1}_{\dot\gamma_1} x^{\alpha_2}_{\dot\gamma_2} x^{\alpha_3}_{\dot\gamma_3} ( \partial^{\dot\gamma_1\gamma_1}\partial^{\dot\gamma_2\gamma_2} \partial^{\dot\gamma_3\gamma_3} ) 
x_{\gamma_1}^{\dot\beta_1}x_{\gamma_2}^{\dot\beta_2} x_{\gamma_3}^{\dot\beta_3} \, {1\over (x^2)^2}\,.
\end{align}
The generalization to an arbitrary number of derivatives is  straightforward
\begin{align}\label{ten-inversion}
 I\big[ \partial^{\dot\alpha_1\beta_1} \ldots \partial^{\dot\alpha_n\beta_n}\big] =  (x^2)^2 
x^{\alpha_1}_{\dot\gamma_1} \ldots x^{\alpha_n}_{\dot\gamma_n} ( \partial^{\dot\gamma_1\gamma_1}\ldots \partial^{\dot\gamma_n\gamma_n} )
x_{\gamma_1}^{\dot\beta_1}\ldots x_{\gamma_n}^{\dot\beta_n} \, {1\over (x^2)^2}\,.
\end{align}
This relation can be proved by induction. Contracting the indices on both sides of \re{ten-inversion} with 
$\epsilon_{\beta_i\beta_{i+1}}$ and using the
identity $\partial^{\dot\alpha_1\beta_1} \partial^{\dot\alpha_2\beta_2}\epsilon_{\beta_1\beta_2}=-\epsilon^{\dot\alpha_1\dot\alpha_2} \Box$,  we obtain
\begin{align}\label{use}
 I\big[ \Box\big]  
=  (x^2)^3    \Box  {1\over x^2}\,,\qquad 
I\big[ \partial^{\dot\alpha_1\beta_1}\Box\big]  
=  (x^2)^3 x^{\alpha_1}_{\dot\gamma_1}  (\partial^{\dot\gamma_1\gamma_1} \Box)  x_{\gamma_1}^{\dot\beta_1} {1\over x^2}
\,,\qquad  I[ \Box^2] =  (x^2)^4 \Box^2\,.
\end{align}
where $\Box=\partial^\mu \partial_\mu$.

\section{$\mathcal N=4$ superconformal generators} \label{App-trick}

In this appendix we elucidate the origin of the maximal number of derivatives recipe that we used in section \ref{sect:trick} to compute 
various components of the super correlation function.

It is  well known that the special superconformal $\bar S-$transformations can be realized as a superposition
of inversion and chiral super-Poincar\'e $Q-$transformations
\begin{align}\label{bar-S}
\bar S_{\dot\alpha}^A  = I \, Q_{\alpha}^A\, I\,, 
\end{align}
with the composite index $A=(a,a')$. The super-Poincar\'e transformations generated by
$Q_{\alpha}^a$ and $Q_\alpha^{a'}$ take the form
\begin{align} \notag\label{Q-tran}
&  x'_{i,\alpha\dot\alpha} =x_{i,\alpha\dot\alpha}+ \epsilon^{a'}_{\alpha} \bar\theta_{i,a'\dot\alpha}\,,&&  
 y'{}_{i,a'}^a=y_{i,a'}^a\,,
  \\[2mm]
 &  \theta'{}_{i,\alpha}^a =\theta_{i,\alpha}^a+\epsilon_{\alpha}^a+ \epsilon^{a'}_{\alpha} y_{i,a'}^a  \,, &&   
  \bar\theta'_{i,a'\dot\alpha}=\bar\theta_{i,a'\dot\alpha}\,,
\end{align}
where $x_i'=e^{\epsilon \cdot Q  +\epsilon' \cdot Q'} x_i $ and similarly for the other coordinates. The action of inversion on the 
supercoordinates looks as
\begin{align}\notag\label{inv-all}
& I[x_i^{\alpha \dot\beta}] = (x_i^{-1})^{\beta\dot\alpha}  \,,&&
I[y_{i a'}{}^a] =  y_{i a'}{}^a + \theta_i^{a\alpha} { (x_i^{-1})_{\alpha\dot\alpha} }\bar\theta_{i,a'}^{\dot\alpha}
\\[2mm] 
& I[\theta^a_{i,\alpha}]= (x_i^{-1})^\beta_{\dot\alpha} \theta^a_{i,\beta}  
\,,&&
I[\bar\theta_{i,a'\dot\alpha}]= - \bar\theta_{i,a' \dot\beta} (x_i^{-1})^{\dot\beta}_{\alpha}\,.
\end{align}
Combining these relations we  obtain from \re{bar-S} the global form of the $\bar S-$transformations 
\begin{align}\notag\label{sbar}
& x_i'' = x_i(1- \bar\xi'\bar\theta_i)^{-1} \,,
& & y_i'' =  (1-\bar\theta_i\bar\xi')^{-1}(y_i+\bar\theta_i \bar\xi)\,,
\\[2mm]
& \bq_i'' = \bar\theta_i(1- \bar\xi'\bar\theta_i)^{-1} \,,
&& \q_i'' =\q_i + x_i 
  (1- \bar\xi'\bar\theta_i)^{-1}(\bar \xi+ \bar\xi' y_i)\,,
\end{align}
where $x_i''=\e^{\,\bar\xi \cdot \bar S  +\bar\xi' \cdot \bar S'} x_i$ and we did not display the Lorentz and $SU(4)$ indices for the sake of simplicity. One can verify that the generators of the transformations \re{sbar} are given by the operators $\bar S_{\dot\alpha}^a$ and $\bar S_{\dot\alpha}^{a'}$ defined in \re{gens}.

The expression for the supercorrelator \re{master} involves the differential operator $\widetilde S_{b'\dot\beta}$ which is related to 
the  generator $\bar S_{b'\dot\beta}$ by a similarity transformation \re{tildeS}. It is easy to see using \re{gens} that 
\begin{align}\notag\label{tilde-S1}
\widetilde S_{b'\dot\beta} =\bar S_{b'\dot\beta}+2\sum_i \bq _{i,b'\dot\beta} {}&= {1\over (x_1^2 x_2^2 x_3^2 x_4^2)^2} \bar S_{b'\dot\beta} (x_1^2 x_2^2 x_3^2 x_4^2)^2
\\
{}& = {1\over (x_1^2 x_2^2 x_3^2 x_4^2)^2} I \, Q_{b'\beta}\, I  (x_1^2 x_2^2 x_3^2 x_4^2)^2\,,
\end{align}
where in the second line we applied \re{bar-S}.  It follows from the first relation that $\{\widetilde S_{b'\dot\beta}, Q_{a'\alpha} \} =0$. Then, we apply \re{tilde-S1} to get
\begin{align}\label{S-Q}
\widetilde S'{}^4 =  {1\over (x_1^2 x_2^2 x_3^2 x_4^2)^2} I \, Q'{}^4\, I  (x_1^2 x_2^2 x_3^2 x_4^2)^2\,,
\end{align}
where $\widetilde S'{}^4=\prod_{b',\dot\beta}\widetilde S_{b'\dot\beta}$ and similarly for $Q'{}^4$.

Let us now examine the action of $Q'{}^4$ on a test function $f(x)$. Replacing the $Q'-$generator by its explicit expression \re{gens} we find
\begin{align}\notag\label{Q-expand}
Q'{}^4 f(x) {}& = \prod_{a', \alpha}\Big[\sum_{1\le i\le 4} \bq_{i,a'\dot\alpha} (\partial_{x_i})^{\dot\alpha\alpha}\Big]  f(x)
\\
{}& =\frac1{12}\sum_{1\le i_1,i_2,i_3,i_4\le 4} (\bq_{i_1}\partial_{i_1})^{\alpha_1a_1'}(\bq_{i_2}\partial_{i_2})_{a_1'}^{\alpha_2}
(\bq_{i_3}\partial_{i_3})^{a_2'}_{\alpha_2}(\bq_{i_4}\partial_{i_4})_{a_2' \alpha_1}f(x)\,,
\end{align}
where we used the notation for $ (\bq_{i}\partial_{i})^{\alpha a'} = (\bq_{i})^{a'}_{\dot\beta} (\partial_{x_i})^{\dot\beta\alpha}$ and the $SU(2)$ indices $a$ and $a'$ are lowered and raised in the same manner as the Lorentz indices, e.g, $\q^a= \epsilon^{ab} \q_b$
and $\bq_{a'} =\epsilon_{a'b'} \bq^{b'}$
 (see Appendix \ref{App-conv}). Each term on the right-hand
side of \re{Q-expand} involves four spatial derivatives and the same number of $\bq-$variables. 

We can use \re{S-Q} together with \re{Q-expand} to obtain an analogous expression for $\widetilde S'{}^4 f(x)$. To this end, we
first apply inversion to both sides of \re{Q-expand}  
\begin{align}
I\, Q'{}^4 f(x) =  I\, Q'{}^4\, I \,(x_1^2 x_2^2 x_3^2 x_4^2)^2 \phi(x) = (x_1^2 x_2^2 x_3^2 x_4^2)^2 \widetilde S'{}^4 \phi(x)\,,
\end{align}
with  $\phi(x) = I\left[(x_1^2 x_2^2 x_3^2 x_4^2)^2f(x) \right]$. To evaluate $I\, Q'{}^4\, I$
we apply the identity 
\begin{align}
  I\big[ (\bq \partial)^{a_1\alpha_1} \ldots (\bq \partial)^{a_n\alpha_n}\big]   =(x^2)^2(\bq \partial)^{a_1\beta_1} \ldots (\bq \partial)^{a_n\beta_n}\,
x_{\beta_1}^{\dot\alpha_1}\ldots x_{\gamma_n}^{\dot\alpha_n}{1\over (x^2)^2 }
\end{align}
that follows from \re{ten-inversion} and \re{inv-all}. In this way, we arrive at
\begin{align}\notag\label{S-magic}
\widetilde S'{}^4 \phi(x) = \frac1{12}\sum_{1\le i_1,i_2,i_3,i_4\le 4} {}& 
 (\bq_{i_1}\partial_{i_1})^{\alpha_1a_1'}(\bq_{i_2}\partial_{i_2})_{a_1'}^{\alpha_2}
(\bq_{i_3}\partial_{i_3})^{\alpha_3a_2'} (\bq_{i_4}\partial_{i_4})_{a_2'}^{\alpha_4}
\\
{}&\quad \times (x_{i_1})_{\alpha_1\dot\alpha_1} (x_{i_2})_{\alpha_2\dot\alpha_2} 
(x_{i_3})_{\alpha_3}^{\dot\alpha_2}(x_{i_4})_ {\alpha_4}^{\dot\alpha_1} \,\phi(x) \,.
\end{align}
Notice that the spatial derivatives in the first line do not commute with the product of $x$'s in the second line.
Let us compare \re{S-magic} with the analogous relation in which we replace $ \widetilde S'$ by its explicit expression \re{gens}
\begin{align}\label{S-naive}
\widetilde S'{}^4 \phi(x) =\prod_{a', \dot\alpha}\Big[\sum_{1\le i\le 4} (\bq_{i}\,\partial_{i})^{a'\alpha} (x_{i})_{\alpha\dot\alpha}+ \bq_{i,\dot\beta}^{\, a'} \bq_{i,b'\dot\alpha}
 (\partial_{\bq_i})^{b'\dot\beta} \Big] \phi(x)\,.
\end{align}
We observe that in order to reproduce \re{S-magic} it is sufficient to neglect the second term inside the  brackets in \re{S-naive}, and then move all spatial derivatives to the left of all $x-$dependent factors. It is this shortcut that we 
used in section \ref{sect:trick}.

\bibliographystyle{JHEP}

\end{document}